\documentclass[11pt,a4paper,twocolumn]{article} 

\usepackage{abstract}
\usepackage{authblk}
\usepackage[utf8]{inputenc}         
\usepackage[T1]{fontenc}            
\usepackage{lmodern}                
\usepackage{upgreek}                
\usepackage{float}
\usepackage{balance}

\usepackage{amsmath,amssymb}        
\usepackage{physics}                

\usepackage{graphicx}               
\usepackage{caption}                
\usepackage{subcaption}             

\usepackage[a4paper,margin=0.75in]{geometry} 
\usepackage{hyperref}               
\hypersetup{                          
  colorlinks=true,                   
  linkcolor=blue,                    
  citecolor=blue,                    
  urlcolor=blue                      
}

\usepackage{ graphicx}
\usepackage{  amsfonts,amsmath,amssymb,amstext}
\usepackage{  float,wrapfig}
\usepackage{  psfrag}
\usepackage{  dsfont}
\usepackage{  color}
\usepackage{  comment}
\usepackage{  bm}
\usepackage{  mathtools}
\definecolor{purple}{rgb}{0.8,0,0.6}
\definecolor{orange}{rgb}{1,0.64,0}

\usepackage[backend=biber,style=nature,sorting=none]{biblatex} 
\numberwithin{equation}{section}    

\begin{document}

\title{\textbf{\Large Magnetoelectric Switching of Magnetic Order in Rhombohedral Graphene}}

\author[1,*]{Kilian Krötzsch}
\author[2]{Arsen Herasymchuk}
\author[1]{Yaroslav Zhumagulov}
\author[3]{Arnaud Magrez}
\author[4]{Kenji Watanabe}
\author[5]{Takashi Taniguchi}
\author[2,6]{Sergei G. Sharapov}
\author[1]{Oleg V. Yazyev}
\author[1,7,$\dagger$]{Mitali Banerjee}
\affil[1]{Institute of Physics, École Polytechnique Fédérale de Lausanne (EPFL), CH-1015 Lausanne, Switzerland}
\affil[2]{Bogolyubov Institute for Theoretical Physics, Kyiv, 03143, Ukraine}
\affil[3]{Crystal Growth Facility (IPHYS-CGCP), Institute of Physics, École Polytechnique Fédérale de Lausanne (EPFL), CH-1015 Lausanne, Switzerland}
\affil[4]{Research Center for Functional Materials, National Institute for Materials Science, 1-1 Namiki, Tsukuba 305-0044, Japan}
\affil[5]{International Center for Materials Nanoarchitectonics,
National Institute for Materials Science, 1-1 Namiki, Tsukuba 305-0044, Japan}
\affil[6]{Kyiv Academic University, 03142 Kyiv, Ukraine}
\affil[7]{Center for Quantum Science and Engineering (QSE Center), École Polytechnique Fédérale de Lausanne (EPFL), CH-1015 Lausanne, Switzerland}
\affil[*]{\textup{kilian.krotzsch@epfl.ch}}
\affil[$\dagger$]{\textup{Corresponding author: mitali.banerjee@epfl.ch}}

\date{} 

\twocolumn[
\begin{@twocolumnfalse}
\maketitle
\vspace*{-2.5em}  

\begin{abstract}
A finite Hall conductance under zero magnetic field implies time reversal symmetry (TRS) breaking due to magnetic order. In rhombohedral stacked multilayer graphene, the angular momentum that breaks TRS can result from the orbital degree of freedom at the $K$ and $K'$ valleys. This leads to valley polarization and occupation-dependent anomalous Hall resistance (AHR) due to the chirality in Berry curvature at the valleys. We report magnetoelectric control of orbital magnetic order in crystalline rhombohedral hexalayer graphene (R6G), achieved without the introduction of a moiré superlattice. At moderate displacement fields and low carrier densities, we observe a non-volatile and hysteretic AHR that can be electrically toggled by sweeping either the carrier density or the displacement field. Upon the application of small perpendicular magnetic fields, the system reveals a characteristic double sign reversal of the AHR, indicating a competition between distinct magnetic ground states. This interplay between valley polarization and electric and magnetic field tuning demonstrates the rich multiferroic behavior of R6G. Our findings present crystalline R6G as a minimal, tunable platform for studying symmetry-breaking phases and magnetic order in flat-band systems, offering insights into the coupling between electronic structure and magnetoelectric response.
\end{abstract}

\vspace{1em} 
\end{@twocolumnfalse}
]    

Multilayer graphene in the rhombohedral (RH) stacking order features a low-energy dispersion following a power law as $E \sim k^N$, where $N$ is the layer number and $k$ is the wave vector. At low energy and charge carrier density, there is a pair of touching conduction and valence bands that become extremely flat, which corresponds to a quenching of the kinetic energy of the electrons~\cite{paul_nery_ab-initio_2021, min_electronic_2008, guinea_electronic_2007, koshino_trigonal_2009, mak_electronic_2010, guinea_electronic_2006}. The resulting high density of states enhances the significance of electron–electron interactions. While an increase in layer number should lead to progressively flatter bands, the trigonal warping effects in the low energy band structure of RH graphene increase band dispersion again~\cite{koshino_trigonal_2009}. The ``sweetspot'' in terms of layer number was calculated to be around $N$ = 5~\cite{han_correlated_2024}. Additionally, the low-energy states in the valleys located at the $K$ and $K'$ points of the Brillouin zone have a large and homogeneous Berry curvature.~\cite{koshino_trigonal_2009, han_orbital_2023, Slizovskiy2019}. In the case of energy degeneracy between the two valleys, their Berry curvatures sum up to zero and ensure that time-reversal symmetry (TRS) is preserved. If the valley degeneracy is spontaneously broken due to electron interactions, the system acquires a net orbital momentum, which is known as the ferro-valleytronic effect~\cite{Zhu2020, Xu2014, Xiao2010, han_orbital_2023}. The finite net orbital momentum breaks TRS and yields an anomalous Hall conductance (AHC), reaching a quantized value if the chemical potential is fully inside the gap in the energy spectrum~\cite{Haldane1988, Chen2020, Polshyn2020}. In this case of complete valley degeneracy lifting, the Berry curvature will integrate to an integer Chern number of $N$ for $N$-layer RH graphene~\cite{koshino_trigonal_2009}. Since the $K$ and $K'$ valleys contribute opposite orbital magnetizations that depend on the absolute sign of an applied electric displacement field, the system yields a valley-magnetic quartet in total~\cite{han_orbital_2023}. 

In this work, we investigate the magnetoelectric switching of competing magnetic orders in crystalline rhombohedral hexalayer graphene (R6G). While previous work on this system (five- and six-layer RH graphene) focused mostly on the low or high electric displacement field regime, our observations were realized at moderate electric displacement fields~\cite{Han2023, han_orbital_2023, https://doi.org/10.48550/arxiv.2503.09954, https://doi.org/10.48550/arxiv.2504.05129}. We report an AHR that is electrically switchable in a non-volatile fashion, exhibiting hysteresis with respect to changes in the carrier density and the application of a sweeping electric displacement field. This is achieved without the need for a stabilizing applied parallel or perpendicular magnetic field as previously observed~\cite{Choi2025,https://doi.org/10.48550/arxiv.2508.15909}. In addition, the magnetic order competition in the system leads to a characteristic double sign change in the AHR upon applying an external magnetic field, alongside clear magnetic hysteresis. This double-switch is dependent on both the electric displacement field, as well as the gate-defined carrier density and its sweeping direction. These results highlight the intricate tunability and sensitivity of the AHC caused by the orbital magnetic order in crystalline R6G.
\begin{figure*}[t]
    \centering
    \includegraphics[width=1\linewidth]{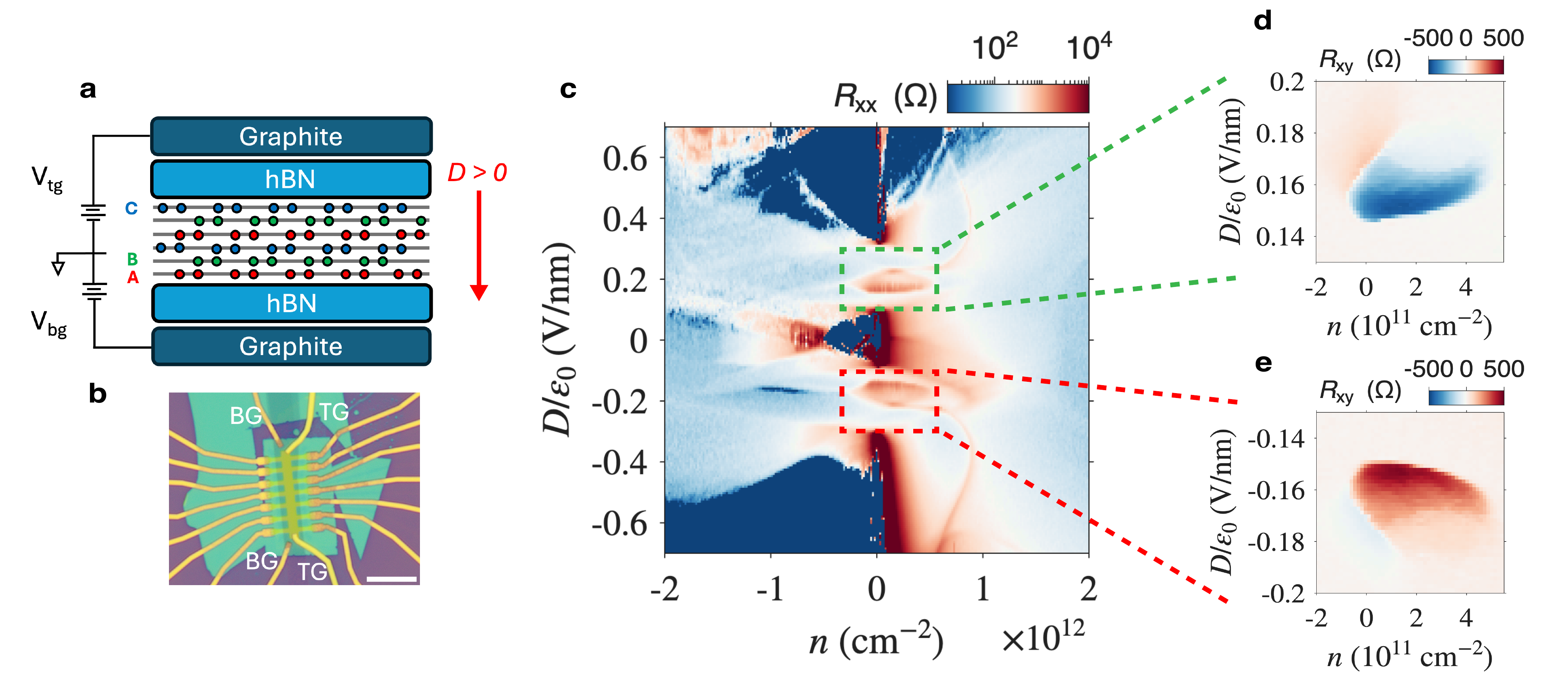} 
    \caption{\textbf{Device structure and electronic transport characteristics.}\\ \textbf{a}, Schematic drawing of the device. R6G is fully encapsulated between two flakes of hBN. Dual graphite gates (top and bottom) enable independent control over the vertical electric displacement field $D$ and carrier density $n$. The displacement field is defined to be positive when directed from the top gate to the bottom gate. \textbf{b}, Optical microscope image of the fabricated device. Electrical contacts to the top gate (TG) and bottom gate (BG) are labeled. Scale bar: 10 $\upmu$m. \textbf{c}, 2D color map of the longitudinal resistance $R_{\text{xx}}$ as a function of carrier density $n$ and displacement field $D$ at zero magnetic field. Two distinct regions of interest characterized by tunable multiferroic behavior emerge at moderate positive and negative values of $D$, and are highlighted in green and red, respectively. \textbf{d}, Zoomed-in 2D color map of the transverse resistance $R_{\text{xy}}$ in the green region as highlighted in panel \textbf{c}. \textbf{e}, Zoomed-in 2D color map of the transverse resistance $R_{\text{xy}}$ in the red region as highlighted in panel \textbf{c}.}
    \label{fig: figure1}
\end{figure*}

\begin{figure*}[t]
    \centering
    \includegraphics[width=1\linewidth]{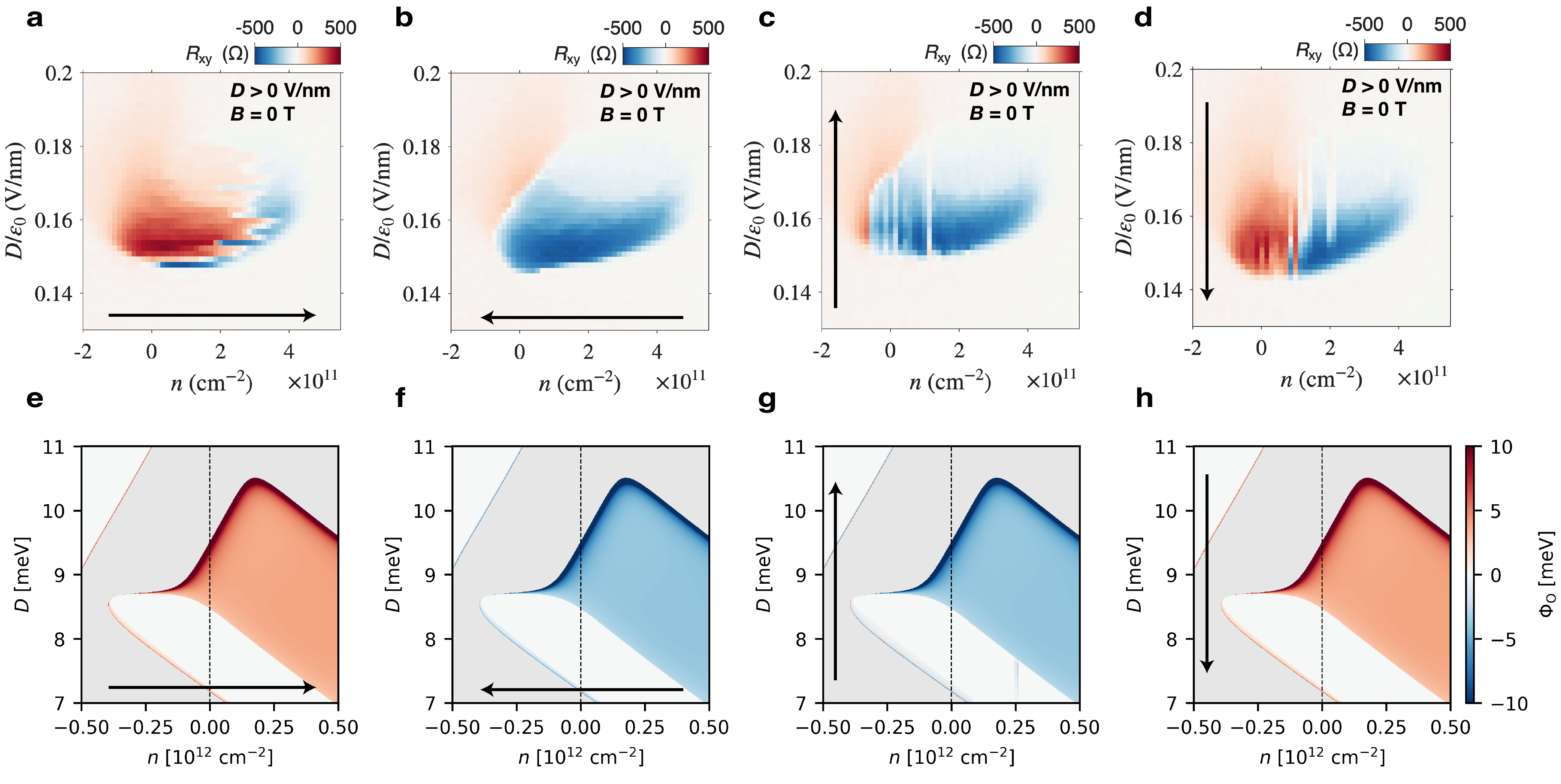} 
    \caption{\textbf{Electric switching of magnetic order and orbital magnetization phase diagram.}\\ \textbf{a--d}, 2D color maps of the transverse resistance $R_{\text{xy}}$ as a function of carrier density $n$ and electric displacement field $D$ at zero magnetic field, shown for the positive displacement field region. Black arrows indicate the direction and axis of the fast parameter sweep. Independent control of the AHR is achieved by sweeping either the carrier density or the displacement field. This tunability is robust against variations in the preparation of the zero-field state and is independent of the direction of the slow sweeping axis. While sweeping the carrier density induces a sign reversal in $R_{\text{xy}}$ over most of the AHR wing, the tunability via the displacement field is more localized near charge neutrality and diminishes at higher doping levels. In the negative $D$ region, the behavior mirrors that of the positive side, with a reversed sign of the transverse resistance. \textbf{e}--\textbf{h}, Calculated phase diagram of the orbital magnetization for R6G in the $n$--$D$-plane. The sign of the orbital magnetization order parameter depends on the direction of the sweeping of the carrier density or the displacement field. Gray areas denote regions where the Landau free-energy model is inapplicable due to a negative quartic coefficient \(c\).}
    \label{fig: figure2}
\end{figure*}

\begin{figure*}[t]
    \centering
    \includegraphics[width=1\linewidth]{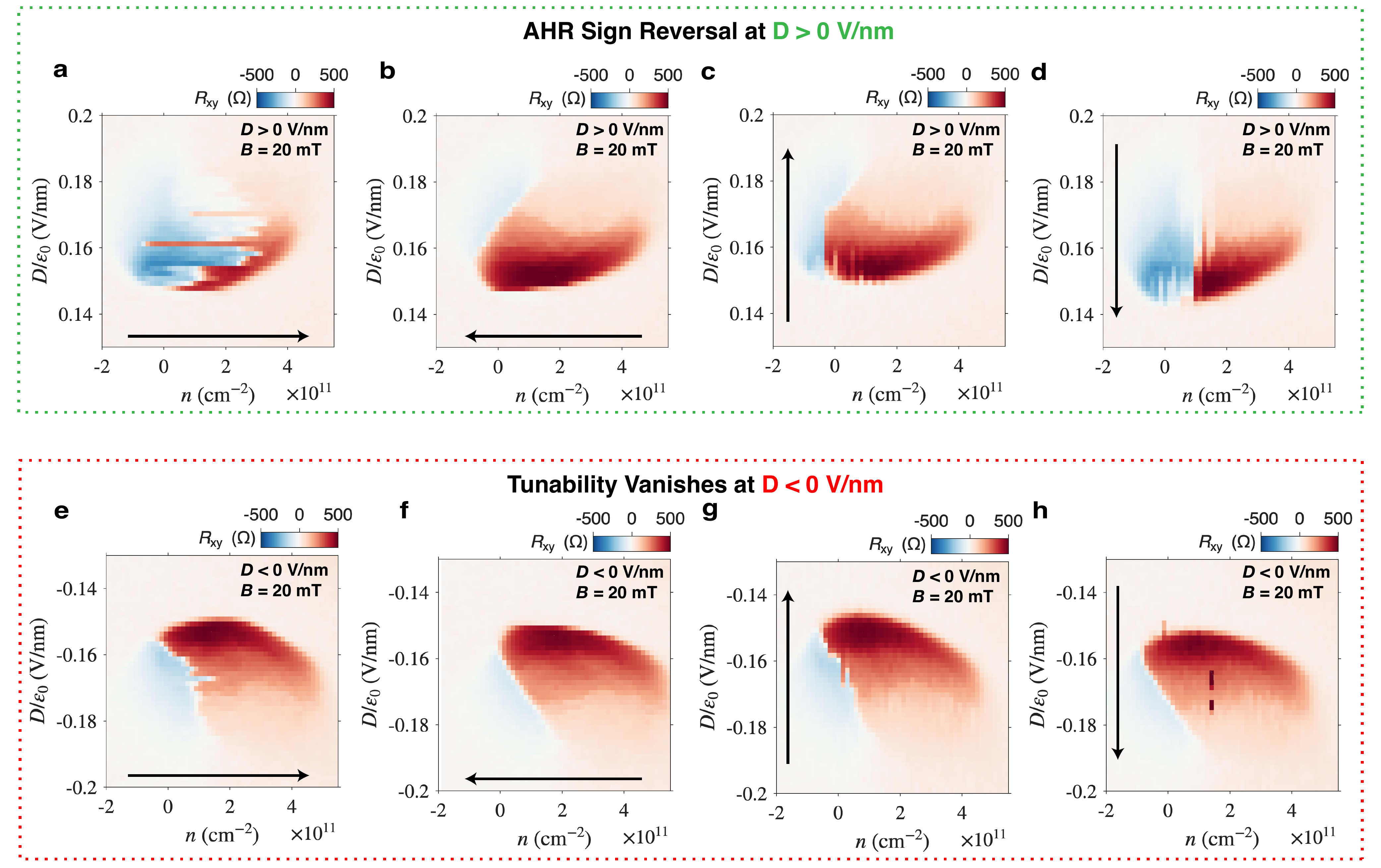} 
    \caption{\textbf{Magnetic-field dependence of electric tunability in the anomalous Hall response.}\\ \textbf{a--d}, 2D color maps of the transverse resistance $R_{\text{xy}}$ as a function of carrier density $n$ and electric displacement field $D$ at a small magnetic field $B = +20$~mT shown for the positive displacement field region. \textbf{e--h}, Equivalent maps for the negative displacement field region. Black arrows indicate the direction and axis of the fast parameter sweep. The tunability of the ground state with carrier density and displacement field persists in the positive displacement field region, albeit with a sign change in the transverse resistance, making it behave like the state in the negative displacement field region at zero magnetic field. The situation is different in the negative displacement field region shown in panels \textbf{e--h}. The AHR remains positive on the electron-doped side, with a sign switch to negative values across the CNP along a diagonal white line in the $n$--$D$-space.}
    \label{fig: figure3}
\end{figure*}

\begin{figure*}[t]
    \centering
    \includegraphics[width=1\linewidth]{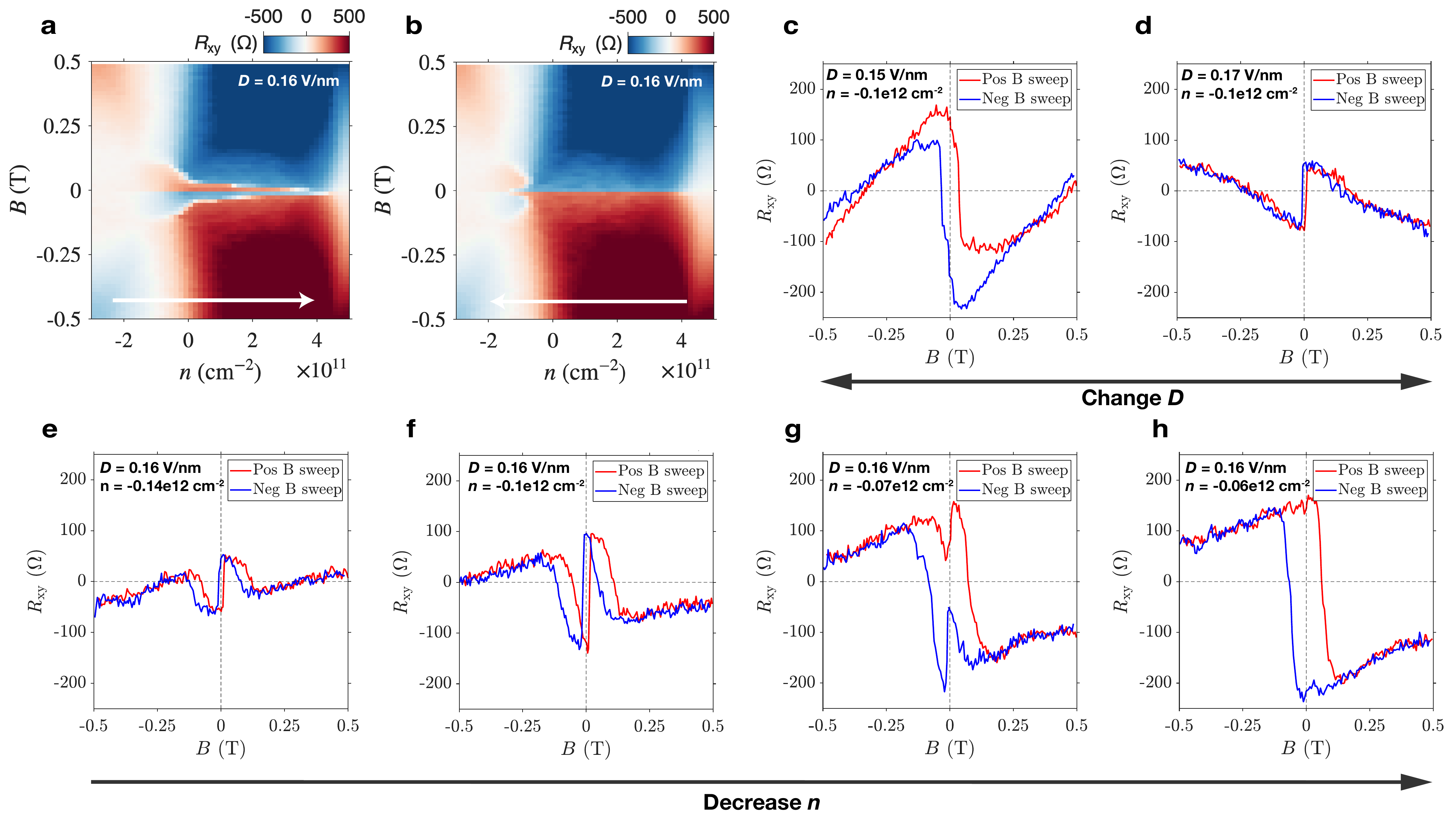} 
    \caption{\textbf{Competition between electrically prepared magnetic order and field-induced AHR sign reversal.}\\ \textbf{a,b}, 2D color maps of the transverse resistance $R_{\text{xy}}$ as a function of carrier density $n$ and perpendicular magnetic field $B$, measured at a fixed displacement field $D/\varepsilon_\text{0}$ = 0.16 V/nm. The carrier density is swept in the direction indicated by the white arrows. As expected, $R_{\text{xy}}$ exhibits a sign change across charge neutrality and upon reversal of the magnetic field. However, an additional spontaneous sign reversal emerges at low magnetic fields, resulting in a characteristic double sign change in $R_{\text{xy}}$. This feature can persist up to higher electron doping when the system is prepared by sweeping the carrier density from hole to electron doping, as evident in the differences between panels \textbf{a} and \textbf{b}. \textbf{c--d}, Transverse resistance $R_{\text{xy}}$ plotted as a function of perpendicular magnetic field for fixed $n=-0.1\times 10^{12}$~cm$^{-2}$ and varying displacement fields. \textbf{e--h}, Transverse resistance $R_{\text{xy}}$ plotted as a function of perpendicular magnetic field for fixed $D/\varepsilon_\text{0}$ = 0.16 V/nm and decreasing carrier densities. Magnetic field sweep directions are indicated in blue (positive-to-negative) and red (negative-to-positive). At specific carrier densities, the hysteresis curves display a double sign reversal in $R_{\text{xy}}$, which evolves into a single, rectangular hysteresis loop at intermediate doping levels.}
    \label{fig: figure4}
\end{figure*}

\textbf{Results}\\
\textbf{Non-volatile electric switching of magnetic order}\\
Device details are illustrated in Figure~\ref{fig: figure1}\textbf{a}. The R6G is encapsulated between layers of hBN, and dual graphite gates are employed above and below the stack. These enable independent control over both the electric displacement field across the R6G and the carrier density within the system. The displacement field is defined to be positive ($D > 0$) when oriented from the top to the bottom gate. An optical microscope image of the device is shown in Figure~\ref{fig: figure1}\textbf{b}. The large RH stacked region allows the fabrication of multiple electrical contacts, enabling a study of transport homogeneity across the sample. In contrast to devices based on twisted graphene layers, where spatial inhomogeneity in twist angle often results in significant variations in physical properties at the micrometer scale, the R6G device shows consistent and reproducible transport behavior across different contact pairs~\cite{Padhi2020, Shen2020, Mukherjee2025}. Figure~\ref{fig: figure1}\textbf{c} displays a 2D color map of the longitudinal resistance $R_{\text{xx}}$ as a function of carrier density $n$ and electric displacement field $D$ at zero magnetic field. The transport characteristics differ significantly from those of Bernal-stacked multilayer graphene. This work focuses on the symmetric regions at moderate positive and negative electric displacement fields, indicated by green and red boxes in Figure~\ref{fig: figure1}\textbf{c}. Zoomed-in maps of the marked regions of interest are shown in Figure~\ref{fig: figure1}\textbf{d} and \textbf{e}. At low carrier densities, the chemical potential lies within the intrinsic low-energy flat bands of R6G. In this regime, the AHR is shown to be magnetoelectrically tunable and enables robust, non-volatile resistance sign switching.

We begin the transport characterization at zero magnetic field to investigate the spontaneous magnetic order, its orientation, and its distribution as a function of the tuned carrier density and applied electric displacement field. We focus on the positive displacement field region, highlighted by the green box in Figure~\ref{fig: figure1}\textbf{c}. Figure~\ref{fig: figure2}\textbf{a--f} present 2D color maps of the transverse resistance $R_{\text{xy}}$ as a function of carrier density $n$ and electric displacement field $D$. The black arrows on the axes denote the fast sweep direction during the measurement. When the carrier density is swept as the fast axis (panels \textbf{a, b}), we observe an almost complete reversal of the AHR sign. However, an intrinsic preference for a particular magnetic ordering persists, resulting in a negative AHR on the electron-doped side at carrier densities around $n \approx 3 \times 10^{11}$ cm$^{-2}$, extending up to charge neutrality for an electric displacement field of approximately $D/\varepsilon_0 \approx 0.145$~V/nm. This non-volatile carrier-density-dependent sign switching of the AHR displays robust tunability at a constant electric displacement field. Since the AHR is not quantized as in the quantum anomalous Hall effect, and the system shows finite longitudinal resistance, the chemical potential is not inside a gap~\cite{Zhu2020, Polshyn2020}. A strict distinction between edge and bulk contribution to the AHR when sweeping the chemical potential $\mu$ directly proportional to $n$ as in the case of the gapped state is thus not applicable here. Such behavior could follow from valley polarization and would challenge conventional understanding~\cite{https://doi.org/10.48550/arxiv.2503.00837}. When we tune the system into the negative displacement field region marked by the red box in Figure~\ref{fig: figure1}\textbf{c}, we observe behavior that is fully analogous but with an opposite sign for the AHR across all combinations of fast axes and sweeping directions. The top row of panel \textbf{e} in Extended Data Figure~\ref{fig: supp3}, where carrier density is the fast axis, demonstrates this flipped but consistent tunability. This is expected due to the electric displacement field sign-dependent Berry curvature and subsequent orbital magnetization at a fixed valley~\cite{han_orbital_2023}.

Next, we sweep the electric displacement field as the fast axis in both directions (Figure~\ref{fig: figure2}\textbf{c, d}). The system exhibits direction-dependent switching of the Hall resistance, but only within a narrow window of low carrier density between $n \approx -0.6$ and $0.92 \times 10^{11}$ cm$^{-2}$. Notably, this tunability is governed purely by the sweep direction, without crossing the zero electric displacement field line. To illustrate this electric switching of the magnetic ground state, line cuts at relevant fixed values of doping and displacement field are shown in Extended Data Figure~\ref{fig: supp7}\textbf{e--f}. Additionally, in both the positive and the negative displacement field regions, the AHR sign at very low carrier densities is negative when the displacement field is being swept from lower to higher values, and vice versa (see Extended Data Figure~\ref{fig: supp3}\textbf{b,~e}). To further verify that the observed tunability is intrinsic and not an artifact of the measurement protocol or state preparation, all eight measurement configurations were repeated after initializing the $B = 0$~T state by sweeping the magnetic field from $B = \pm 100$~mT to 0~T. This value is well above the coercive field of $\pm 50$~mT. Additionally, the four configurations in the positive displacement field region were repeated with the slow sweep direction reversed. Neither the method of initializing the $B$ = 0 T state nor the direction of the slow axis sweep had any measurable impact on the observed transport behavior, confirming the robustness and reproducibility of the electric tunability in both the positive and negative displacement field regimes (see Extended Data Figures~\ref{fig: supp4} and  \ref{fig: supp5}).

From our theoretical analysis, switching can arise from the interplay of two symmetry-breaking orders: valley polarization (orbital magnetization) and spin–valley polarization. We compute the orbital magnetization phase diagram, shown in Figure~\ref{fig: figure2}\textbf{e--h}, by finding the minimum of the Landau free energy derived from an effective tight-binding Hamiltonian for R6G with short-range Coulomb interactions in a two-valley model~(see supplementary note).
The Landau free energy density $f(\Phi)$ is given by
\begin{equation}\tag{1}
\small
\begin{aligned}
f(\Phi)&= a\bigg[ \Phi_{\text{O}}^2  +\sum_{\xi=\pm}  \Phi_{\text{S},\,\xi}^2\bigg]+b\sum_{\xi=\pm} \xi  \Phi_{\text{O}} \Phi_{\text{S},\,\xi}^2 \\
&+c\bigg[  \Phi_{\text{O}}^4 + \sum_{\xi=\pm} 2    \Phi_{\text{S},\,\xi}^4+ \sum_{\xi=\pm} 6 \Phi_{\text{O}}^2 \Phi_{\text{S},\,\xi}^2 \bigg],
\label{eq:landau}
\end{aligned} 
\end{equation}
where \(a\), \(b\), and \(c\) are the Landau parameters that depend on external electric field, doping level, and interaction strengths~(see supplementary note). The order parameters \(\Phi_{\mathrm{O}}\) and \(\Phi_{\mathrm{S},\,\xi}\) denote the orbital magnetization and the spin–valley magnetization in $K$ ($\xi=+$) and $K^{\prime}$($\xi=-$) valleys, respectively. The cubic coupling \(\propto \Phi_{\mathrm{O}}\Phi_{\mathrm{S},\,\xi}^2\), together with the biquadratic term \(\propto \Phi_{\mathrm{O}}^2\Phi_{\mathrm{S},\,\xi}^2\), enforces a first-order transition, producing a sharp, discontinuous change between phases with opposite valley polarization depending on the sweeping direction of the parameters $n$ and $D$, as demonstrated in Figure~\ref{fig: figure2}\textbf{e--h}. The spin and orbital magnetizations modify the quasiparticle dispersion and the associated Berry curvature~\cite{Auerbach2025}. The sign of the Berry curvature depends on the sign of the orbital magnetization parameter \(\Phi_{\text{O}}\) as was mentioned in~\cite{Auerbach2025, Slizovskiy2019}.
The correspondence between the sign of the AHR (Figure~\ref{fig: figure2}\textbf{a–d}) and that of $\Phi_{\text{O}}$ (Figure~\ref{fig: figure2}\textbf{e–h}) demonstrates the direct link between these two quantities. In addition, because the Berry curvature is odd under reversal of the displacement field \(D\), the AHR changes sign under \(D\!\to\!-D\)~\cite{han_orbital_2023}, which explains AHR sign reversal, presented in Extended Data Figure~\ref{fig: supp3}\textbf{b},\textbf{e}. 
Therefore, the observed sign of the AHC results from the interplay between the displacement-field-induced and the interaction-induced Berry curvature modification. Finally, we also observe different types of order mixing in R6G. When the orbital magnetization order parameter is nonzero, the orbital and spin magnetizations become mixed in both valleys. In the white region, where $\Phi_{\text{O}} = 0$, we obtain that only the spin magnetization order in the valleys with $\Phi_{\text{S},+} = \Phi_{\text{S},-}$ exists; see Extended Data Figures~\ref{fig: supp8} and \ref{fig: supp9}.

\textbf{Reversal and suppression of electric field tunability}\\
The system exhibits a more intricate behavior when a finite perpendicular magnetic field is applied. In the positive displacement field region, the Hall resistance remains tunable in a non-volatile fashion through both the direction of the gate-induced carrier density sweep and variations in the electric displacement field. If a negative magnetic field of $B = -20$~mT is applied, the system's behavior is entirely unchanged. However, if a positive magnetic field of $B = +20$~mT is applied, the sign of the transverse resistance is reversed, and the system behaves similarly to the negative displacement field region at zero magnetic field (Figure~\ref{fig: figure3}\textbf{a–d}). Conversely, when a small negative magnetic field of $B = -20$~mT is turned on, the Hall resistance in the negative displacement field region undergoes a sign reversal, making it resemble the behavior of the positive displacement field region at zero field. Notably, both $B = \pm 20$~mT remain below the coercive field threshold of $B = \pm 50$~mT.

In contrast, in the negative displacement field region under a positive magnetic field of $B~=~+20$~mT, the tunability of the AHR for both carrier density and displacement field is entirely suppressed (Figure~\ref{fig: figure3}\textbf{e–h}). This suppression is robust and reproducible between measurements, and shows no dependence on the scan direction of the slow axis. The AHR wing merely undergoes a sign change from positive to negative resistance when crossing a diagonal white zero resistance line in the $n$--$D$-space close to the charge neutrality point (CNP). The external perpendicular magnetic field lifts the valley energy degeneracy, resulting in a finite valley splitting due to the distinct orbital magnetic moments of the valleys. This explicitly induced valley polarization can enhance or invert the spontaneous orbital magnetization up to a critical magnetic field at which the bistability is suppressed. It would then appear natural to assume that the tunability of the AHR wing in the $D<0$ and $B>0$ case should remain unchanged. However, the effect of an external magnetic field on this system is nontrivial, and its influence can be qualitatively evaluated by incorporating it into the derived Landau free energy of Eq.~(\ref{eq:landau}) as follows:
\begin{equation}\tag{2}
f_{\mathrm{mag}}(\Phi)
= 2g_{\mathrm{O}}\,\Phi_{\mathrm{O}} B_{z}
+ g_{\mathrm{S}}\bigl(\Phi_{\mathrm{S},+}-\Phi_{\mathrm{S},-}\bigr)\,|\mathbf{B}|,
\label{eq:fmag}
\end{equation}
where $g_{\mathrm{O}}$ and $g_{\mathrm{S}}$ are the orbital and spin $g$ factors, respectively. Because $g_{\mathrm{O}}$ greatly exceeds $g_{\mathrm{S}}$ in the $z$-direction~\cite{Xiao2007}, the first term dominates for an out-of-plane field. Moreover, the sign of $g_{\mathrm{O}}$ is secured by the sign of the displacement field $D$~\cite{Xiao2007}. Taken together, these facts imply that $\Phi_{\mathrm{O}}$ can be tuned nontrivially by $D$ and $B$. In particular, AHR switching is obtained only for positive product of $D$ and $B$; otherwise, the AHR sign is fixed by the lowering Landau free energy with the term $g_{\mathrm{O}}\,\Phi_{\mathrm{O}} B_{z}$ (see Extended Data Fig.~\ref{fig: supp3}).

\textbf{Competing magnetic order and hysteretic behavior}\\
Finally, we investigate the system’s response to a sweeping external perpendicular magnetic field to probe the competing magnetic orders at small fields as observed in hBN-moiré systems at one charge carrier per moiré unit cell~\cite{https://doi.org/10.48550/arxiv.2506.01485, https://doi.org/10.48550/arxiv.2503.00837,https://doi.org/10.48550/arxiv.2405.16944,Ding2025,https://doi.org/10.48550/arxiv.2412.09985}. Figure~\ref{fig: figure4}\textbf{a–b} present 2D color maps of the transverse resistance $R_{\text{xy}}$ as a function of carrier density $n$ and magnetic field $B$, at a fixed electric displacement field of $D/\varepsilon_0 = 0.16$ V/nm. As expected, the Hall resistance changes sign upon crossing the CNP, which corresponds to a transition from hole to electron carriers, and upon reversal of the magnetic field. However, an additional sign change in $R_{\text{xy}}$ is observed that persists up to $B = \pm 100$~mT in the slightly hole-doped regime, around $n = -1~\times~10^{11}$~cm$^{-2}$. This feature extends to higher electron doping levels when the carrier density is swept from the hole-doped to the electron-doped side, as shown in Fig.~\ref{fig: figure4}\textbf{a}, albeit up to lower magnetic fields. Conversely, if the carrier density is swept in the opposite direction, this double sign reversal of the transverse resistance is absent at higher electron densities. To further explore this behavior, we examined magnetic hysteresis at various fixed displacement fields and carrier densities. Representative hysteresis loops, measured at different carrier densities and fixed electric displacement fields, are shown in Fig.~\ref{fig: figure4}\textbf{e–h}. In these measurements, the carrier density was swept on the hole-doped side. Beyond a critical carrier density, or if the electric displacement field is changed (Fig.~\ref{fig: figure4}\textbf{c,d}), the characteristic double switching of the Hall resistance vanishes. Instead, conventional magnetic hysteresis emerges, indicative of magnetic order that aligns with the external magnetic field.

Small external magnetic fields will introduce a finite energy splitting between the valleys, leading to valley polarization. The Berry curvature of the occupied valley determines the orbital magnetization. When the magnetic field crosses zero, valley polarization and occupation are flipped, which causes a reversal of the Hall resistance. If the orbital magnetization is antiparallel to the magnetically induced Hall resistance, a sufficiently large field will suppress the spontaneous ordering, leading to the observed additional sign switch. However, if the system is prepared by sweeping the carrier density in the opposite direction, as shown in Figure~\ref{fig: figure4}\textbf{b}, the orbital magnetization is already aligned with the external magnetic field, and the double sign switch vanishes, showing only conventional hysteresis.

\textbf{Conclusion}\\
We have demonstrated robust switching of the anomalous Hall resistance arising from orbital magnetization in crystalline rhombohedral hexalayer graphene at low carrier densities and under moderate electric displacement fields. The observed magnetic ordering can be reversibly tuned at zero magnetic field by varying the sweeping direction of either (i) the induced carrier density or (ii) the electric displacement field. Furthermore, when both the carrier density and displacement field are fixed, the competition between magnetic orders can be continuously tuned through (iii) the application of a small perpendicular magnetic field. The direction-dependent switching behavior is well captured by Landau free-energy minimization with microscopic parameterization, which predicts first-order phase transitions between states of opposite valley polarization. This results in a bistable magnetic system that exhibits hysteretic reversals in the sign of the anomalous Hall resistance. Our findings enrich the phase diagram of crystalline multilayer rhombohedral graphene by demonstrating non-volatile switching between competing ground states driven by magnetoelectric coupling. They uncover the intricate interplay between the spin and valley degrees of freedom in correlated two-dimensional systems.

\textbf{Acknowledgments}\\
K.K. acknowledges very helpful discussions with Zhengguang Lu and aid in device fabrication from Ó. Manuel Rios Alves. K.K. acknowledges funding from SNSF. M.B. acknowledges the support of the SNSF Eccellenza grant No. PCEGP2\_194528, and support from the QuantERA II Programme that has received funding from the European Union’s Horizon 2020 research and innovation program under Grant Agreement No 101017733. K.W. and T.T. acknowledge support from the JSPS KAKENHI (Grant Numbers 20H00354 and 23H02052) and World Premier International Research Center Initiative (WPI), MEXT, Japan. A.H., Y.Z., S.G.S., and O.V.Y. acknowledge support from the SNSF through the Ukrainian-Swiss Joint research project ``Transport and thermodynamic phenomena in low-dimensional materials with flat bands'' (grant No. IZURZ2\_224624).

\textbf{Author contributions}\\
M.B. and K.K. conceived the project. M.B. supervised the project. K.K. fabricated the devices, performed the measurements, and analyzed the data with input from M.B. A.H., Y.Z. carried out the theory work under the guidance of S.G.S. and O.V.Y. K.W. and T.T. provided the hBN crystals. A.M. provided the Raman spectrometer. K.K. wrote the manuscript with inputs from all the authors. 

\textbf{Data availability}\\
The data supporting the findings of this study are available from the corresponding author upon reasonable request.

\appendix
\textbf{Methods}\\
\textbf{Device fabrication}\\
Graphite/graphene and hBN were first exfoliated onto an O$_2$-plasma cleaned SiO$_2$-coated (285 nm) Si-substrate, followed by detailed optical microscope characterization. Suitable homogeneous hBN-flakes for encapsulation ($\approx$ 30 nm) and graphite-flakes for gating ($\approx$ 2–3 nm) were selected and their quality confirmed via AFM imaging. The layer number of the hexalayer graphene flake was determined via calibrated optical contrast, and the RH domain was resolved via Raman spectroscopy mapping of the 2D vibrational interlayer mode. The selected RH domain was isolated via anodic oxidation in an AFM using a conductive tip to reduce the chance of relaxation into a Bernal stacking order during fabrication (see Extended Data Figure~\ref{fig: supp1}). The bottom graphite gate and bottom hBN layer were picked up employing an all dry-transfer method by using a polypropylene carbonate (PPC) covered PDMS stamp in an Argon-filled glovebox. The bottom part of the material stack was annealed for one hour at 350°C in forming gas. The top graphite gate, top hBN layer, and the hexalayer graphene were picked up with a poly(bisphenol A-carbonate) (PC) covered PDMS stamp and released onto the prepared bottom part. The device structure was created with standard electron beam lithography and reactive ion etching techniques. The electrical connections were established with e-beam evaporation of Cr/Au (5~nm/70~nm).

\textbf{Electrical transport measurements}\\
The measurements were carried out in two separate dilution refrigerators with base phonon temperatures of 8–12 mK. The longitudinal and transverse resistances were measured with Zurich Instruments MFLI lock-in amplifiers at an excitation current of 0.5–5 nA demodulated at a frequency of either 7.777 Hz or 17.777 Hz. The gate voltages were applied using Yokogawa GS200 and Keithley 2400 source-meters. The carrier density and displacement field were determined according to $n=(C_\text{TG}V_\text{TG}+C_\text{BG}V_\text{BG})/e$ and $D/\varepsilon_0=(C_\text{TG}V_\text{TG}-C_\text{BG}V_\text{BG})/2\varepsilon_0$, respectively, where $C_\text{TG/BG}$ and $V_\text{TG/BG}$ are the top and bottom gate capacitances per unit area and voltages, respectively. The gate capacitances depend on the hBN layer thicknesses and were determined via Hall resistance measurements at different gate-defined carrier densities. The AHR measurements were immediately acquired in the $n$--$D$-space with defined fast sweeping axes and directions. To obtain the hysteretic measurements, the fast axis was swept in both directions before incrementing the slow axis value.
\balance
\newpage
\printbibliography[heading=none]

@article{paul_nery_ab-initio_2021,
	title = {Ab-initio energetics of graphite and multilayer graphene: stability of Bernal versus rhombohedral stacking},
	volume = {8},
	issn = {2053-1583},
	url = {https://dx.doi.org/10.1088/2053-1583/abec23},
	doi = {10.1088/2053-1583/abec23},
	shorttitle = {Ab-initio energetics of graphite and multilayer graphene},
	pages = {035006},
	number = {3},
	journaltitle = {2D Materials},
	shortjournal = {2D Mater.},
	author = {Paul Nery, Jean and Calandra, Matteo and Mauri, Francesco},
	urldate = {2025-01-30},
	date = {2021-03},
	langid = {english},
	note = {Publisher: {IOP} Publishing},
}

@article{min_electronic_2008,
	title = {Electronic Structure of Multilayer Graphene},
	volume = {176},
	issn = {0375-9687},
	url = {https://doi.org/10.1143/PTPS.176.227},
	doi = {10.1143/PTPS.176.227},
	pages = {227--252},
	journaltitle = {Progress of Theoretical Physics Supplement},
	shortjournal = {Progress of Theoretical Physics Supplement},
	author = {Min, Hongki and {MacDonald}, Allan H.},
	urldate = {2024-08-14},
	date = {2008-06-01},
}

@article{guinea_electronic_2007,
	title = {Electronic properties of stacks of graphene layers},
	volume = {143},
	issn = {0038-1098},
	url = {https://www.sciencedirect.com/science/article/pii/S0038109807002931},
	doi = {10.1016/j.ssc.2007.03.053},
	series = {Exploring graphene},
	pages = {116--122},
	number = {1},
	journaltitle = {Solid State Communications},
	shortjournal = {Solid State Communications},
	author = {Guinea, F. and Castro Neto, A. H. and Peres, N. M. R.},
	urldate = {2025-01-30},
	date = {2007-07-01},
}

@article{koshino_trigonal_2009,
	title = {Trigonal warping and Berry's phase \$N{\textbackslash}ensuremath\{{\textbackslash}pi\}\$ in {ABC}-stacked multilayer graphene},
	volume = {80},
	url = {https://link.aps.org/doi/10.1103/PhysRevB.80.165409},
	doi = {10.1103/PhysRevB.80.165409},
	pages = {165409},
	number = {16},
	journaltitle = {Physical Review B},
	shortjournal = {Phys. Rev. B},
	author = {Koshino, Mikito and {McCann}, Edward},
	urldate = {2024-07-01},
	date = {2009-10-12},
	note = {Publisher: American Physical Society},
}

@article{mak_electronic_2010,
	title = {Electronic Structure of Few-Layer Graphene: Experimental Demonstration of Strong Dependence on Stacking Sequence},
	volume = {104},
	url = {https://link.aps.org/doi/10.1103/PhysRevLett.104.176404},
	doi = {10.1103/PhysRevLett.104.176404},
	shorttitle = {Electronic Structure of Few-Layer Graphene},
	pages = {176404},
	number = {17},
	journaltitle = {Physical Review Letters},
	shortjournal = {Phys. Rev. Lett.},
	author = {Mak, Kin Fai and Shan, Jie and Heinz, Tony F.},
	urldate = {2025-01-30},
	date = {2010-04-29},
	note = {Publisher: American Physical Society},
}

@article{guinea_electronic_2006,
	title = {Electronic states and Landau levels in graphene stacks},
	volume = {73},
	url = {https://link.aps.org/doi/10.1103/PhysRevB.73.245426},
	doi = {10.1103/PhysRevB.73.245426},
	pages = {245426},
	number = {24},
	journaltitle = {Physical Review B},
	shortjournal = {Phys. Rev. B},
	author = {Guinea, F. and Castro Neto, A. H. and Peres, N. M. R.},
	urldate = {2025-01-30},
	date = {2006-06-21},
	note = {Publisher: American Physical Society},
}

@article{han_correlated_2024,
	title = {Correlated insulator and Chern insulators in pentalayer rhombohedral-stacked graphene},
	volume = {19},
	rights = {2023 The Author(s), under exclusive licence to Springer Nature Limited},
	issn = {1748-3395},
	url = {https://www.nature.com/articles/s41565-023-01520-1},
	doi = {10.1038/s41565-023-01520-1},
	pages = {181--187},
	number = {2},
	journaltitle = {Nature Nanotechnology},
	shortjournal = {Nat. Nanotechnol.},
	author = {Han, Tonghang and Lu, Zhengguang and Scuri, Giovanni and Sung, Jiho and Wang, Jue and Han, Tianyi and Watanabe, Kenji and Taniguchi, Takashi and Park, Hongkun and Ju, Long},
	urldate = {2024-08-09},
	date = {2024-02},
	langid = {english},
	note = {Publisher: Nature Publishing Group},
}

@article{han_orbital_2023,
	title = {Orbital multiferroicity in pentalayer rhombohedral graphene},
	volume = {623},
	rights = {2023 The Author(s), under exclusive licence to Springer Nature Limited},
	issn = {1476-4687},
	url = {https://www.nature.com/articles/s41586-023-06572-w},
	doi = {10.1038/s41586-023-06572-w},
	pages = {41--47},
	number = {7985},
	journaltitle = {Nature},
	author = {Han, Tonghang and Lu, Zhengguang and Scuri, Giovanni and Sung, Jiho and Wang, Jue and Han, Tianyi and Watanabe, Kenji and Taniguchi, Takashi and Fu, Liang and Park, Hongkun and Ju, Long},
	urldate = {2024-06-20},
	date = {2023-11},
	langid = {english},
	note = {Publisher: Nature Publishing Group},
}

@article{Zhu2020,
  title = {Voltage-Controlled Magnetic Reversal in Orbital Chern Insulators},
  volume = {125},
  ISSN = {1079-7114},
  url = {http://dx.doi.org/10.1103/PhysRevLett.125.227702},
  DOI = {10.1103/physrevlett.125.227702},
  number = {22},
  journal = {Physical Review Letters},
  publisher = {American Physical Society (APS)},
  author = {Zhu,  Jihang and Su,  Jung-Jung and MacDonald,  A. H.},
  year = {2020},
  month = nov 
}

@misc{https://doi.org/10.48550/arxiv.2503.00837,
  doi = {10.48550/ARXIV.2503.00837},
  url = {https://arxiv.org/abs/2503.00837},
  author = {Wang,  Zhiyu and Liu,  Qianling and Han,  Xiangyan and Li,  Zhuoxian and Zhao,  Wenjun and Qu,  Zhuangzhuang and Han,  Chunrui and Watanabe,  Kenji and Taniguchi,  Takashi and Han,  Zheng Vitto and Zhou,  Sicheng and Tong,  Bingbing and Liu,  Guangtong and Lu,  Li and Liu,  Jianpeng and Wu,  Fengcheng and Lu,  Jianming},
  title = {Electrical switching of Chern insulators in moire rhombohedral heptalayer graphene},
  publisher = {arXiv},
  year = {2025},
  copyright = {Creative Commons Attribution Non Commercial Share Alike 4.0 International}
}

@misc{https://doi.org/10.48550/arxiv.2506.01485,
  doi = {10.48550/ARXIV.2506.01485},
  url = {https://arxiv.org/abs/2506.01485},
  author = {Xie,  Jian and Zhang,  Zaizhe and Chen,  Xi and Kwan,  Yves H. and Huo,  Zihao and Herzog-Arbeitman,  Jonah and Guo,  Liangliang and Watanabe,  Kenji and Taniguchi,  Takashi and Liu,  Kaihui and Xie,  X. C. and Bernevig,  B. Andrei and Song,  Zhi-Da and Lu,  Xiaobo},
  title = {Unconventional Orbital Magnetism in Graphene-based Fractional Chern Insulators},
  publisher = {arXiv},
  year = {2025},
  copyright = {Creative Commons Attribution Non Commercial Share Alike 4.0 International}
}

@misc{https://doi.org/10.48550/arxiv.2405.16944,
  doi = {10.48550/ARXIV.2405.16944},
  url = {https://arxiv.org/abs/2405.16944},
  author = {Xie,  Jian and Huo,  Zihao and Lu,  Xin and Feng,  Zuo and Zhang,  Zaizhe and Wang,  Wenxuan and Yang,  Qiu and Watanabe,  Kenji and Taniguchi,  Takashi and Liu,  Kaihui and Song,  Zhida and Xie,  X. C. and Liu,  Jianpeng and Lu,  Xiaobo},
  title = {Tunable Fractional Chern Insulators in Rhombohedral Graphene Superlattices},
  publisher = {arXiv},
  year = {2024},
  copyright = {Creative Commons Attribution Non Commercial Share Alike 4.0 International}
}

@article{Ding2025,
  title = {Electric-Field Switchable Chirality in Rhombohedral Graphene Chern Insulators Stabilized by Tungsten Diselenide},
  volume = {15},
  ISSN = {2160-3308},
  url = {http://dx.doi.org/10.1103/PhysRevX.15.011052},
  DOI = {10.1103/physrevx.15.011052},
  number = {1},
  journal = {Physical Review X},
  publisher = {American Physical Society (APS)},
  author = {Ding,  Jing and Xiang,  Hanxiao and Hua,  Jiannan and Zhou,  Wenqiang and Liu,  Naitian and Zhang,  Le and Xin,  Na and Wu,  Bing and Watanabe,  Kenji and Taniguchi,  Takashi and Sofer,  Zdeněk and Zhu,  Wei and Xu,  Shuigang},
  year = {2025},
  month = mar 
}

@misc{https://doi.org/10.48550/arxiv.2412.09985,
  doi = {10.48550/ARXIV.2412.09985},
  url = {https://arxiv.org/abs/2412.09985},
  author = {Zheng,  Jian and Wu,  Size and Liu,  Kai and Lyu,  Bosai and Liu,  Shuhan and Sha,  Yating and Li,  Zhengxian and Watanabe,  Kenji and Taniguchi,  Takashi and Jia,  Jinfeng and Shi,  Zhiwen and Chen,  Guorui},
  title = {Switchable Chern insulator,  isospin competitions and charge density waves in rhombohedral graphene moire superlattices},
  publisher = {arXiv},
  year = {2024},
  copyright = {Creative Commons Attribution 4.0 International}
}

@article{Choi2025,
  title = {Superconductivity and quantized anomalous Hall effect in rhombohedral graphene},
  volume = {639},
  ISSN = {1476-4687},
  url = {http://dx.doi.org/10.1038/s41586-025-08621-y},
  DOI = {10.1038/s41586-025-08621-y},
  number = {8054},
  journal = {Nature},
  publisher = {Springer Science and Business Media LLC},
  author = {Choi,  Youngjoon and Choi,  Ysun and Valentini,  Marco and Patterson,  Caitlin L. and Holleis,  Ludwig F. W. and Sheekey,  Owen I. and Stoyanov,  Hari and Cheng,  Xiang and Taniguchi,  Takashi and Watanabe,  Kenji and Young,  Andrea F.},
  year = {2025},
  month = mar,
  pages = {342–347}
}

@article{Xu2014,
  title = {Spin and pseudospins in layered transition metal dichalcogenides},
  volume = {10},
  ISSN = {1745-2481},
  url = {http://dx.doi.org/10.1038/nphys2942},
  DOI = {10.1038/nphys2942},
  number = {5},
  journal = {Nature Physics},
  publisher = {Springer Science and Business Media LLC},
  author = {Xu,  Xiaodong and Yao,  Wang and Xiao,  Di and Heinz,  Tony F.},
  year = {2014},
  month = apr,
  pages = {343–350}
}

@article{Xiao2010,
  title = {Berry phase effects on electronic properties},
  volume = {82},
  ISSN = {1539-0756},
  url = {http://dx.doi.org/10.1103/RevModPhys.82.1959},
  DOI = {10.1103/revmodphys.82.1959},
  number = {3},
  journal = {Reviews of Modern Physics},
  publisher = {American Physical Society (APS)},
  author = {Xiao,  Di and Chang,  Ming-Che and Niu,  Qian},
  year = {2010},
  month = jul,
  pages = {1959–2007}
}

@misc{https://doi.org/10.48550/arxiv.2508.15909,
  doi = {10.48550/ARXIV.2508.15909},
  url = {https://arxiv.org/abs/2508.15909},
  author = {Deng,  Jinghao and Xie,  Jiabin and Li,  Hongyuan and Taniguchi,  Takashi and Watanabe,  Kenji and Shan,  Jie and Mak,  Kin Fai and Liu,  Xiaomeng},
  title = {Superconductivity and Ferroelectric Orbital Magnetism in Semimetallic Rhombohedral Hexalayer Graphene},
  publisher = {arXiv},
  year = {2025},
  copyright = {Creative Commons Attribution 4.0 International}
}

@misc{https://doi.org/10.48550/arxiv.2504.05129,
  doi = {10.48550/ARXIV.2504.05129},
  url = {https://arxiv.org/abs/2504.05129},
  author = {Morissette,  Erin and Qin,  Peiyu and Wu,  Hai-Tian and Zhang,  Naiyuan J. and Nguyen,  Ron Q. and Watanabe,  K. and Taniguchi,  T. and Li,  J. I. A.},
  title = {Striped Superconductor in Rhombohedral Hexalayer Graphene},
  publisher = {arXiv},
  year = {2025},
  copyright = {arXiv.org perpetual,  non-exclusive license}
}

@article{Polshyn2020,
  title = {Electrical switching of magnetic order in an orbital Chern insulator},
  volume = {588},
  ISSN = {1476-4687},
  url = {http://dx.doi.org/10.1038/s41586-020-2963-8},
  DOI = {10.1038/s41586-020-2963-8},
  number = {7836},
  journal = {Nature},
  publisher = {Springer Science and Business Media LLC},
  author = {Polshyn,  H. and Zhu,  J. and Kumar,  M. A. and Zhang,  Y. and Yang,  F. and Tschirhart,  C. L. and Serlin,  M. and Watanabe,  K. and Taniguchi,  T. and MacDonald,  A. H. and Young,  A. F.},
  year = {2020},
  month = nov,
  pages = {66–70}
}

@misc{https://doi.org/10.48550/arxiv.2503.09954,
  doi = {10.48550/ARXIV.2503.09954},
  url = {https://arxiv.org/abs/2503.09954},
  author = {Morissette,  Erin and Qin,  Peiyu and Watanabe,  K. and Taniguchi,  T. and Li,  J. I. A.},
  title = {Coulomb-driven Momentum Space Condensation in Rhombohedral Hexalayer Graphene},
  publisher = {arXiv},
  year = {2025},
  copyright = {arXiv.org perpetual,  non-exclusive license}
}

@article{Mukherjee2025,
  title = {Superconducting magic-angle twisted trilayer graphene with competing magnetic order and moiré inhomogeneities},
  ISSN = {1476-4660},
  url = {http://dx.doi.org/10.1038/s41563-025-02252-4},
  DOI = {10.1038/s41563-025-02252-4},
  journal = {Nature Materials},
  publisher = {Springer Science and Business Media LLC},
  author = {Mukherjee,  Ayshi and Layek,  Surat and Sinha,  Subhajit and Kundu,  Ritajit and Marchawala,  Alisha H. and Hingankar,  Mahesh and Sarkar,  Joydip and Sangani,  L. D. Varma and Agarwal,  Heena and Ghosh,  Sanat and Tazi,  Aya Batoul and Watanabe,  Kenji and Taniguchi,  Takashi and Pasupathy,  Abhay N. and Kundu,  Arijit and Deshmukh,  Mandar M.},
  year = {2025},
  month = may 
}

@article{Padhi2020,
  title = {Transport across twist angle domains in moiré graphene},
  volume = {2},
  ISSN = {2643-1564},
  url = {http://dx.doi.org/10.1103/PhysRevResearch.2.033458},
  DOI = {10.1103/physrevresearch.2.033458},
  number = {3},
  journal = {Physical Review Research},
  publisher = {American Physical Society (APS)},
  author = {Padhi,  Bikash and Tiwari,  Apoorv and Neupert,  Titus and Ryu,  Shinsei},
  year = {2020},
  month = sep 
}

@article{Shen2020,
  title = {Correlated states in twisted double bilayer graphene},
  volume = {16},
  ISSN = {1745-2481},
  url = {http://dx.doi.org/10.1038/s41567-020-0825-9},
  DOI = {10.1038/s41567-020-0825-9},
  number = {5},
  journal = {Nature Physics},
  publisher = {Springer Science and Business Media LLC},
  author = {Shen,  Cheng and Chu,  Yanbang and Wu,  QuanSheng and Li,  Na and Wang,  Shuopei and Zhao,  Yanchong and Tang,  Jian and Liu,  Jieying and Tian,  Jinpeng and Watanabe,  Kenji and Taniguchi,  Takashi and Yang,  Rong and Meng,  Zi Yang and Shi,  Dongxia and Yazyev,  Oleg V. and Zhang,  Guangyu},
  year = {2020},
  month = mar,
  pages = {520–525}
}

@article{Han2023,
  title = {Correlated insulator and Chern insulators in pentalayer rhombohedral-stacked graphene},
  volume = {19},
  ISSN = {1748-3395},
  url = {http://dx.doi.org/10.1038/s41565-023-01520-1},
  DOI = {10.1038/s41565-023-01520-1},
  number = {2},
  journal = {Nature Nanotechnology},
  publisher = {Springer Science and Business Media LLC},
  author = {Han,  Tonghang and Lu,  Zhengguang and Scuri,  Giovanni and Sung,  Jiho and Wang,  Jue and Han,  Tianyi and Watanabe,  Kenji and Taniguchi,  Takashi and Park,  Hongkun and Ju,  Long},
  year = {2023},
  month = oct,
  pages = {181–187}
}

@article{PhysRevLett.132.186401,
  title = {Emergent Correlated Phases in Rhombohedral Trilayer Graphene Induced by Proximity Spin-Orbit and Exchange Coupling},
  volume = {132},
  ISSN = {1079-7114},
  url = {http://dx.doi.org/10.1103/PhysRevLett.132.186401},
  DOI = {10.1103/physrevlett.132.186401},
  number = {18},
  journal = {Physical Review Letters},
  publisher = {American Physical Society (APS)},
  author = {Zhumagulov,  Yaroslav and Kochan,  Denis and Fabian,  Jaroslav},
  year = {2024},
  month = may 
}

@article{PhysRevB.110.045427,
  title = {Swapping exchange and spin-orbit induced correlated phases in proximitized Bernal bilayer graphene},
  volume = {110},
  ISSN = {2469-9969},
  url = {http://dx.doi.org/10.1103/PhysRevB.110.045427},
  DOI = {10.1103/physrevb.110.045427},
  number = {4},
  journal = {Physical Review B},
  publisher = {American Physical Society (APS)},
  author = {Zhumagulov,  Yaroslav and Kochan,  Denis and Fabian,  Jaroslav},
  year = {2024},
  month = jul 
}

@article{PhysRevB.14.1165,
  title = {Quantum critical phenomena},
  volume = {14},
  ISSN = {0556-2805},
  url = {http://dx.doi.org/10.1103/PhysRevB.14.1165},
  DOI = {10.1103/physrevb.14.1165},
  number = {3},
  journal = {Physical Review B},
  publisher = {American Physical Society (APS)},
  author = {Hertz,  John A.},
  year = {1976},
  month = aug,
  pages = {1165–1184}
}

@article{PhysRevB.48.7183,
  title = {Effect of a nonzero temperature on quantum critical points in itinerant fermion systems},
  volume = {48},
  ISSN = {1095-3795},
  url = {http://dx.doi.org/10.1103/PhysRevB.48.7183},
  DOI = {10.1103/physrevb.48.7183},
  number = {10},
  journal = {Physical Review B},
  publisher = {American Physical Society (APS)},
  author = {Millis,  A. J.},
  year = {1993},
  month = sep,
  pages = {7183–7196}
}

@article{e2002-00356-9,
  title = {Effective action approach to the Leggett’s mode in two-band superconductors},
  volume = {30},
  ISSN = {1434-6036},
  url = {http://dx.doi.org/10.1140/epjb/e2002-00356-9},
  DOI = {10.1140/epjb/e2002-00356-9},
  number = {1},
  journal = {The European Physical Journal B},
  publisher = {Springer Science and Business Media LLC},
  author = {Sharapov,  S.G. and Gusynin,  V.P. and Beck,  H.},
  year = {2002},
  month = nov,
  pages = {45–51}
}

@article{PhysRevB.82.035409,
  title = {Band structure of ABC-stacked graphene trilayers},
  volume = {82},
  ISSN = {1550-235X},
  url = {http://dx.doi.org/10.1103/PhysRevB.82.035409},
  DOI = {10.1103/physrevb.82.035409},
  number = {3},
  journal = {Physical Review B},
  publisher = {American Physical Society (APS)},
  author = {Zhang,  Fan and Sahu,  Bhagawan and Min,  Hongki and MacDonald,  A. H.},
  year = {2010},
  month = jul 
}

@article{Slizovskiy2019,
  title = {Films of rhombohedral graphite as two-dimensional topological semimetals},
  volume = {2},
  ISSN = {2399-3650},
  url = {http://dx.doi.org/10.1038/s42005-019-0268-8},
  DOI = {10.1038/s42005-019-0268-8},
  number = {1},
  journal = {Communications Physics},
  publisher = {Springer Science and Business Media LLC},
  author = {Slizovskiy,  Sergey and McCann,  Edward and Koshino,  Mikito and Fal’ko,  Vladimir I.},
  year = {2019},
  month = dec 
}

@article{Haldane1988,
  title = {Model for a Quantum Hall Effect without Landau Levels: Condensed-Matter Realization of the “Parity Anomaly”},
  volume = {61},
  ISSN = {0031-9007},
  url = {http://dx.doi.org/10.1103/PhysRevLett.61.2015},
  DOI = {10.1103/physrevlett.61.2015},
  number = {18},
  journal = {Physical Review Letters},
  publisher = {American Physical Society (APS)},
  author = {Haldane,  F. D. M.},
  year = {1988},
  month = oct,
  pages = {2015–2018}
}

@article{Chen2020,
  title = {Tunable correlated Chern insulator and ferromagnetism in a moiré superlattice},
  volume = {579},
  ISSN = {1476-4687},
  url = {http://dx.doi.org/10.1038/s41586-020-2049-7},
  DOI = {10.1038/s41586-020-2049-7},
  number = {7797},
  journal = {Nature},
  publisher = {Springer Science and Business Media LLC},
  author = {Chen,  Guorui and Sharpe,  Aaron L. and Fox,  Eli J. and Zhang,  Ya-Hui and Wang,  Shaoxin and Jiang,  Lili and Lyu,  Bosai and Li,  Hongyuan and Watanabe,  Kenji and Taniguchi,  Takashi and Shi,  Zhiwen and Senthil,  T. and Goldhaber-Gordon,  David and Zhang,  Yuanbo and Wang,  Feng},
  year = {2020},
  month = mar,
  pages = {56–61}
}

@article{Auerbach2025,
  title = {Isospin magnetic texture and intervalley exchange interaction in rhombohedral tetralayer graphene},
  ISSN = {1745-2481},
  url = {http://dx.doi.org/10.1038/s41567-025-03035-z},
  DOI = {10.1038/s41567-025-03035-z},
  journal = {Nature Physics},
  publisher = {Springer Science and Business Media LLC},
  author = {Auerbach,  Nadav and Dutta,  Surajit and Uzan,  Matan and Vituri,  Yaar and Zhou,  Yaozhang and Meltzer,  Alexander Y. and Grover,  Sameer and Holder,  Tobias and Emanuel,  Peleg and Huber,  Martin E. and Myasoedov,  Yuri and Watanabe,  Kenji and Taniguchi,  Takashi and Oreg,  Yuval and Berg,  Erez and Zeldov,  Eli},
  year = {2025},
  month = oct 
}

@article{Xiao2007,
  title = {Valley-Contrasting Physics in Graphene: Magnetic Moment and Topological Transport},
  volume = {99},
  ISSN = {1079-7114},
  url = {http://dx.doi.org/10.1103/PhysRevLett.99.236809},
  DOI = {10.1103/physrevlett.99.236809},
  number = {23},
  journal = {Physical Review Letters},
  publisher = {American Physical Society (APS)},
  author = {Xiao,  Di and Yao,  Wang and Niu,  Qian},
  year = {2007},
  month = dec 
}

\newpage
\clearpage
\setcounter{figure}{0} 
\renewcommand{\figurename}{Extended Data Figure}

\begin{figure*}[t]
    \centering
    \includegraphics[width=1\linewidth]{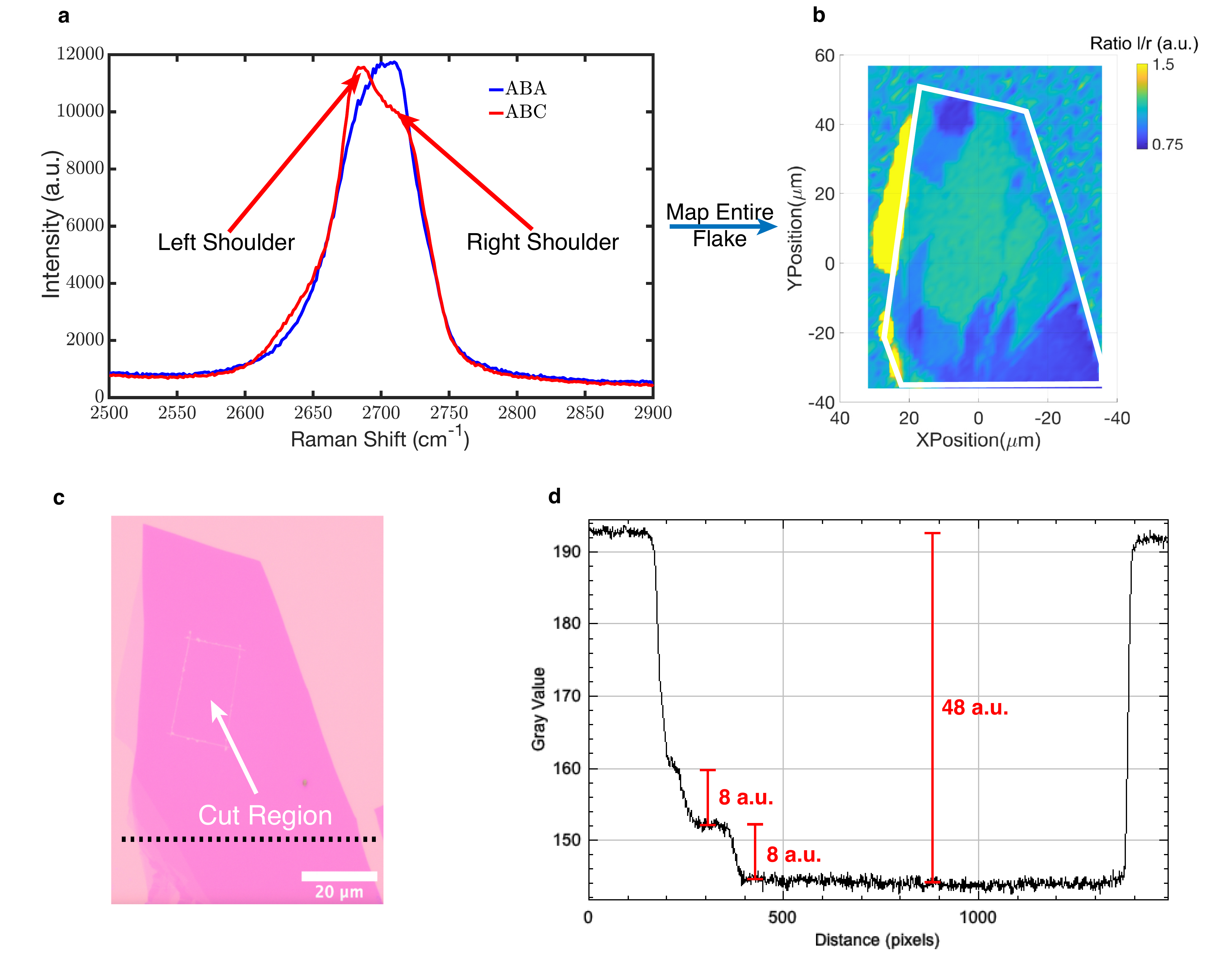} 
    \caption{\textbf{a}, Plot of the 2D interlayer vibrational Raman mode that can be used to identify regions with RH stacking ordering. The ratio of the left-to-right shoulder peak is greater than unity for RH stacking (red line) and smaller than unity for Bernal stacking (blue line). \textbf{b}, The stacking order of the entire flake can be mapped by scanning the Laser along the surface. \textbf{c}, A sufficiently large region of the flake with RH ordering is cut out via anodic oxidation using an AFM. The isolation helps to increase the likelihood that the region remains in RH stacking during fabrication. \textbf{d}, The intensity in arbitrary gray value units (a.u.) plotted against the distance in pixels along the black dashed line in panel \textbf{c} for the green color channel of the optical microscope CCD camera. The optical microscope settings are calibrated against AFM measurements so that 8 a.u. correspond to a single layer of graphene. The 48 a.u. across the graphene multilayer flake corresponds to six layers.}
    \label{fig: supp1}
\end{figure*}

\begin{figure*}[t]
    \centering
    \includegraphics[width=1\linewidth]{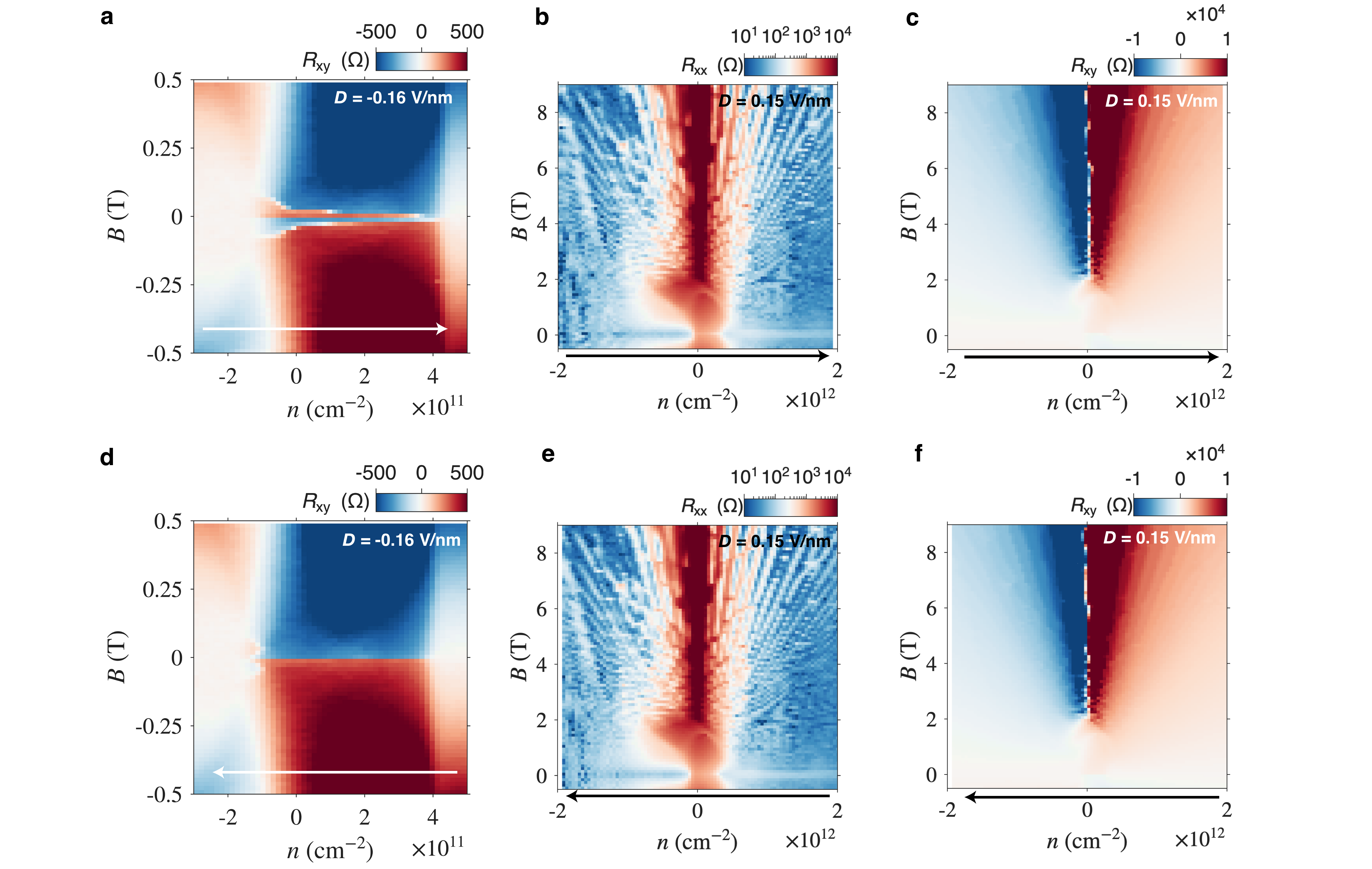} 
    \caption{\textbf{a}, 2D color map of the transverse resistance $R_{\text{xy}}$ as a function of carrier density $n$ and perpendicular magnetic field $B$, measured at a fixed displacement field $D/\varepsilon_\text{0} = -0.16$~V/nm. The carrier density is swept in the direction indicated by the white arrow. \textbf{b, c}, 2D color maps of the longitudinal and transverse resistance $R_{\text{xx}}$ and $R_{\text{xy}}$, respectively, as a function of carrier density $n$ and perpendicular magnetic field $B$, measured at a fixed displacement field $D/\varepsilon_\text{0}$ = 0.15 V/nm. The carrier density is swept in the direction indicated by the black arrows. \textbf{d--f}, The same as in panels \textbf{a--c}, but with the reverse sweeping direction of the fast axis. Panels \textbf{a} and \textbf{d} show analogous behavior of the system as in Figure~\ref{fig: figure4}\textbf{a, b} when exposed to a negative displacement field. The positions of the Dirac point in panels \textbf{b, c, e}, and \textbf{f} were used to correct for the sweeping direction-dependent small absolute offset in the carrier density during the hysteretic measurements.}
    \label{fig: supp2}
\end{figure*}

\begin{figure*}[t]
    \centering
    \includegraphics[width=1\linewidth]{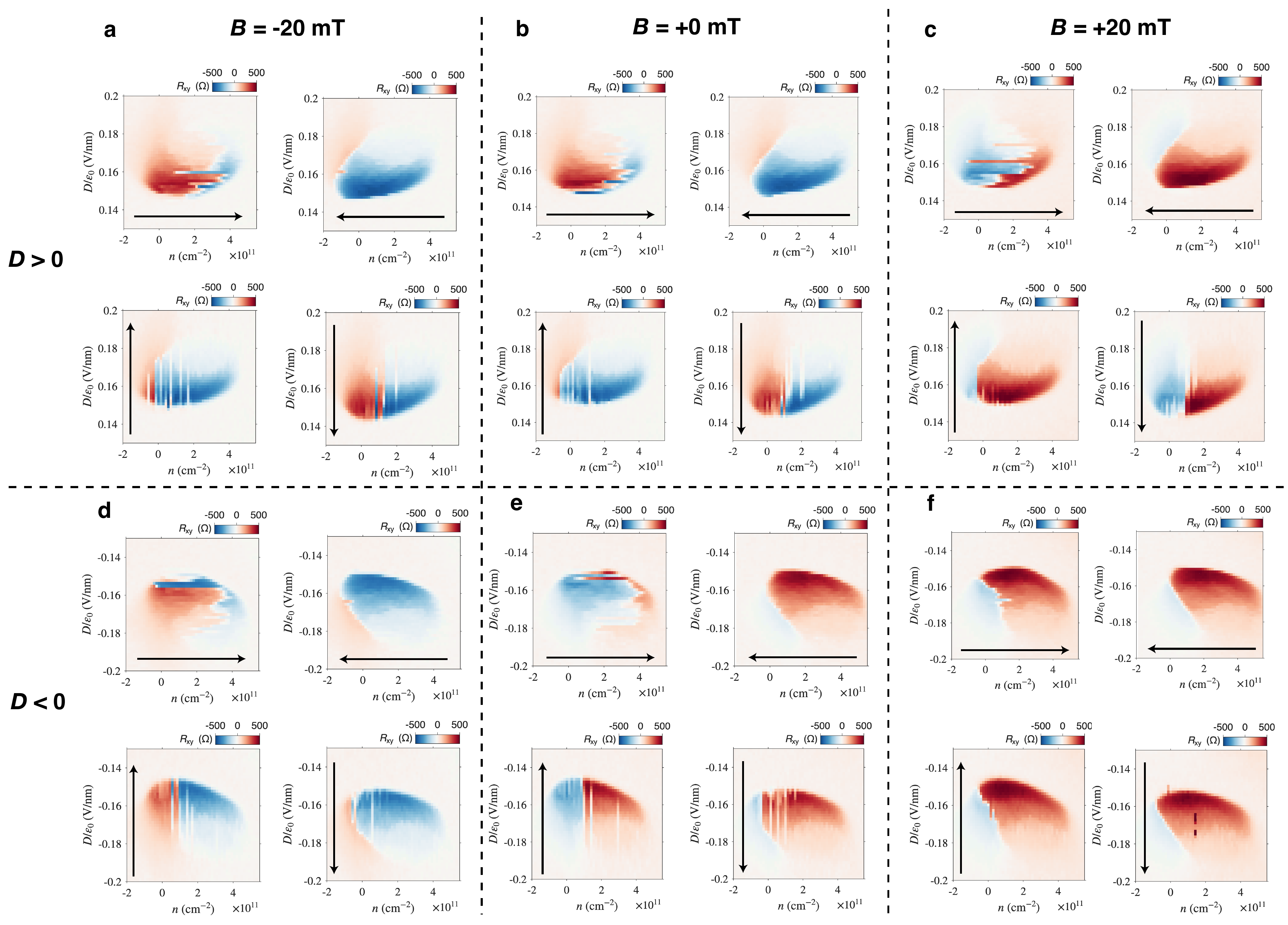} 
    \caption{2D color maps of the transverse resistance $R_{\text{xy}}$ as a function of carrier density $n$ and electric displacement field $D$ for all measured combinations of applied magnetic and displacement fields, and fast sweeping axes. \textbf{a--c}, AHR wing acquired in the positive displacement field region for $B = -$20,0, and $+20$~mT, respectively. \textbf{d--f}, The same as in panels \textbf{a--c}, but in the negative displacement field region.}
    \label{fig: supp3}
\end{figure*}

\begin{figure*}[t]
    \centering
    \includegraphics[width=1\linewidth]{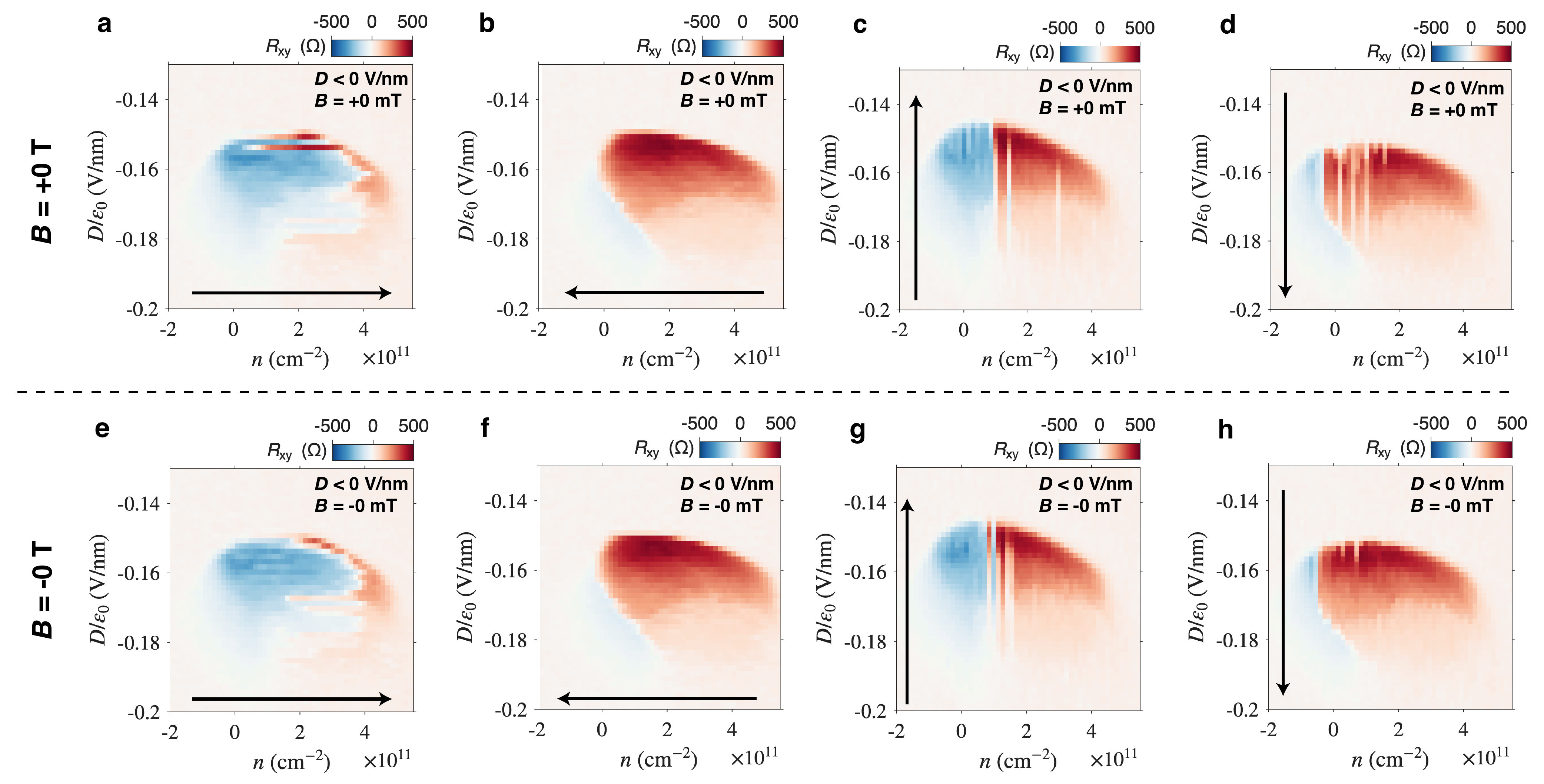} 
    \caption{2D color maps of the transverse resistance $R_{\text{xy}}$ as a function of carrier density $n$ and electric displacement field $D$ at zero magnetic field. The fast sweeping axis is denoted by black arrows. \textbf{a--d}, The system is prepared by increasing the magnetic field above the coercive field of $B=+50$~mT to $B=+100$~mT before ramping it down to zero again. \textbf{e--h}, The same measurements as in panels \textbf{a--d}, but the system was prepared by setting the magnetic field to $B=-100$~mT before ramping it down to zero again. The preparation of the zero magnetic field state does not influence the observed behavior of the system.}
    \label{fig: supp4}
\end{figure*}

\begin{figure*}[t]
    \centering
    \includegraphics[width=1\linewidth]{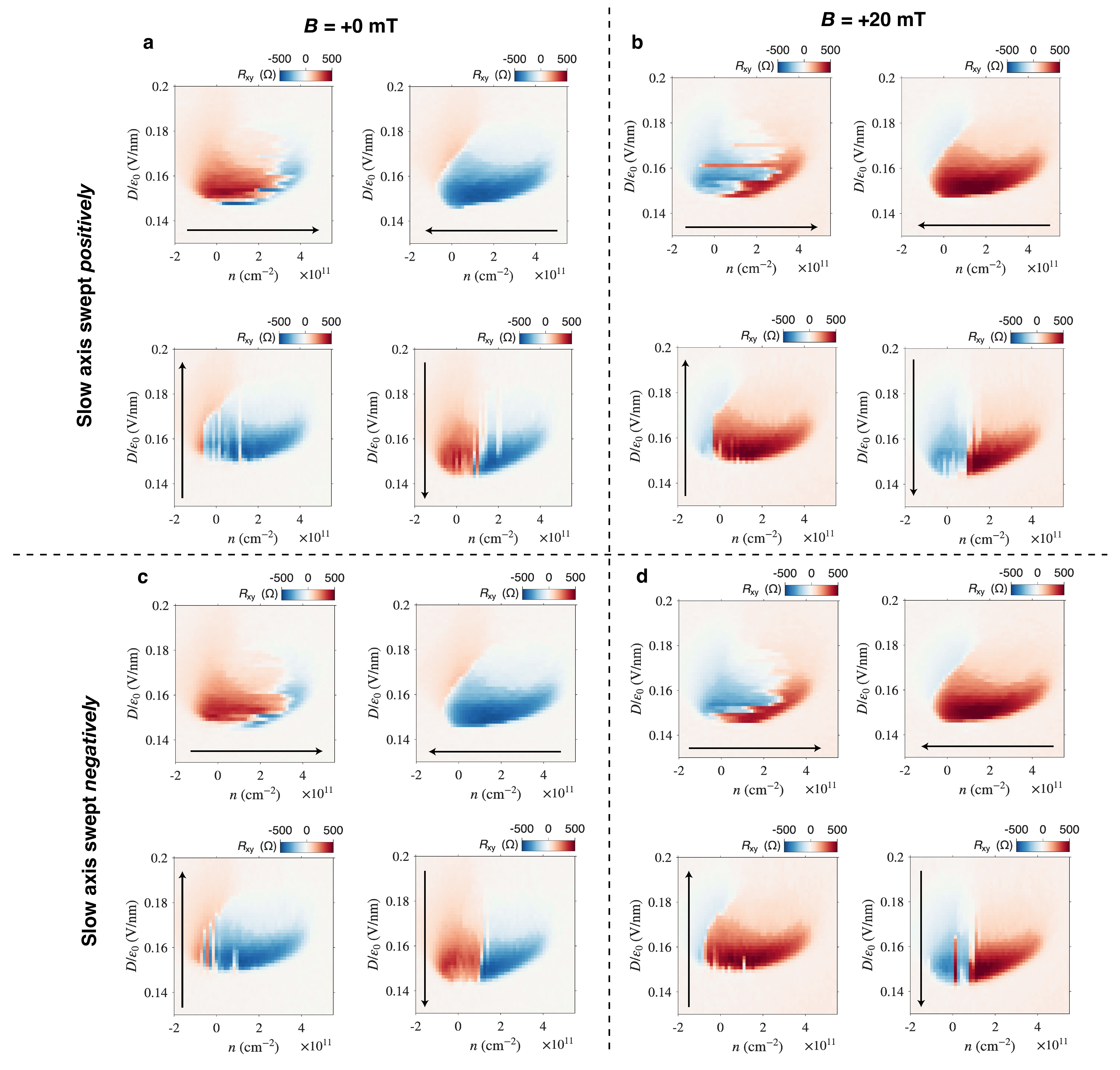} 
    \caption{2D color maps of the transverse resistance $R_{\text{xy}}$ as a function of carrier density $n$ and electric displacement field $D$ at zero and finite magnetic field. The fast sweeping axis is denoted by black arrows. \textbf{a, b}, The slow axis of the respective measurement is swept in the positive direction. \textbf{c, d}, The slow axis of the respective measurement is swept in the negative direction. The sweeping direction of the slow axis has no influence on the observed behavior of the system.}
    \label{fig: supp5}
\end{figure*}

\begin{figure*}[t]
    \centering
    \includegraphics[width=1\linewidth]{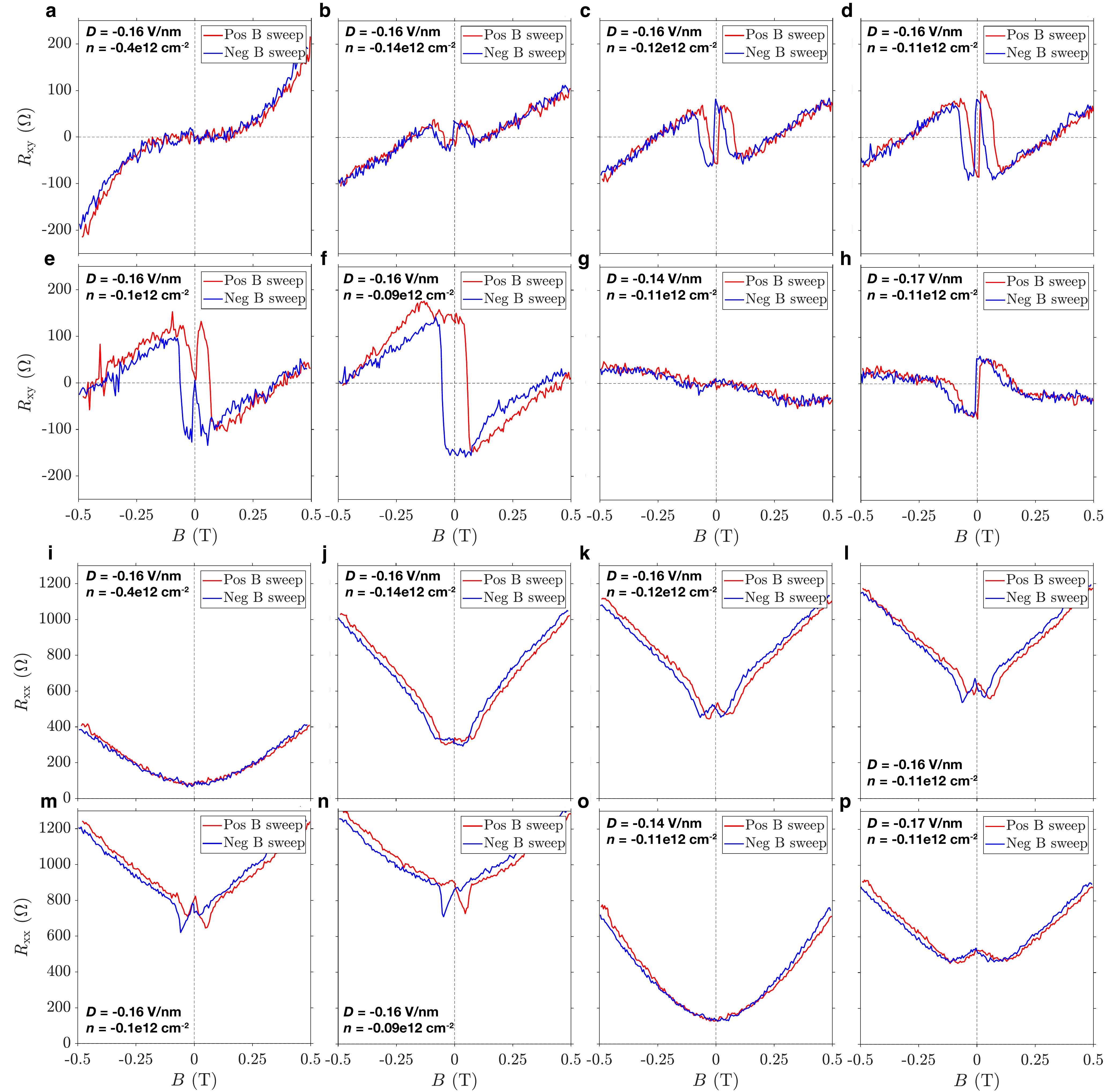} 
    \caption{\textbf{a--h} Transverse resistance $R_{\text{xy}}$ plotted as a function of perpendicular magnetic field for fixed $D/\varepsilon_\text{0} = -0.16$~V/nm and decreasing carrier densities. \textbf{i--p}, The longitudinal resistance $R_{\text{xx}}$ corresponding to the transverse resistance $R_{\text{xy}}$ in panels \textbf{a--h}. The system shows analogous behavior of an emerging double AHR sign switch that develops into normal hysteretic behavior upon changing the carrier density when exposed to negative displacement fields (compare to Figure~\ref{fig: figure4}\textbf{c,d} and \textbf{e--h}).}
    \label{fig: supp6}
\end{figure*}

\begin{figure*}[t]
    \centering
    \includegraphics[width=1\linewidth]{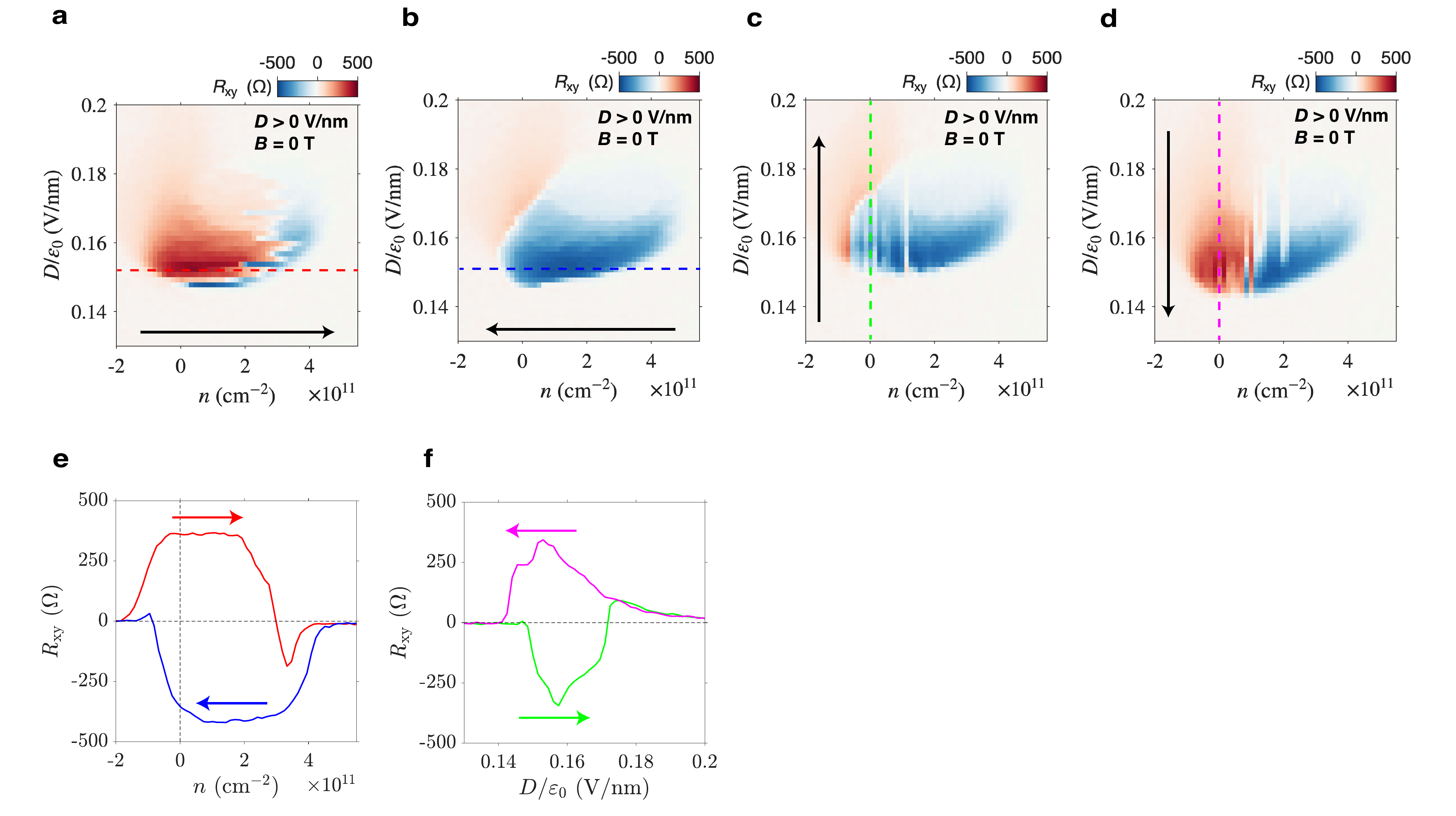} 
    \caption{\textbf{a--d}, 2D color maps of the transverse resistance $R_{\text{xy}}$ as a function of carrier density $n$ and electric displacement field $D$ at zero magnetic field, shown for the positive displacement field region. Black arrows indicate the direction and axis of the fast parameter sweep. \textbf{e--f}, Line cuts taken along the dashed lines in panels \textbf{a--d}, showing $R_{\text{xy}}$ at fixed displacement field and fixed carrier density, respectively. Colored arrows indicate the direction of the fast sweep axis.}
    \label{fig: supp7}
\end{figure*}

\pagebreak

\onecolumn

\section*{Supplementary note on theory}
\label{sec:theoretical-supp}
\renewcommand{\theequation}{S\arabic{equation}}

Throughout this supplementary note, we use $\hbar=c=k_{\mathrm{B}}=1$.

To describe the hysteresis behavior, we use the Landau theory. We consider a system with the following model Hamiltonian density:
\begin{equation}
\begin{aligned}
H(\tau,\mathbf{x})=H_{0}(\tau,\mathbf{x})+H_{\text{int}}(\tau,\mathbf{x}),
\end{aligned} 
\end{equation}
where a free part of the Hamiltonian density is given by
\begin{equation}
\begin{aligned}
H_{0} (\tau,\mathbf{x}) &=\sum_{s ,\xi} \sum_{i,j} \psi^{\dagger}_{i,s \xi} (\tau,\mathbf{x} )\left[ \mathcal{H}_0 (\mathbf{x}; s, \xi )\right]^{ij} \psi_{j,s \xi} (\tau,\mathbf{x} ) \\
\end{aligned} 
\end{equation}
and
\begin{equation}
\label{eq:int-hamiltonian}
\begin{aligned}
H_{\text{int}} (\tau,\mathbf{x}) &=V_0 \left[ n_{\uparrow +} (\tau,\mathbf{x}) n_{\downarrow +} (\tau,\mathbf{x})+ n_{\uparrow -} (\tau,\mathbf{x}) n_{\downarrow -} (\tau,\mathbf{x}) \right]\\
&+V_1 \left[ n_{\uparrow +}(\tau,\mathbf{x})+n_{\downarrow +} (\tau,\mathbf{x}) \right] \left[ n_{\uparrow -} (\tau,\mathbf{x}) +n_{\downarrow -} (\tau,\mathbf{x}) \right],
\end{aligned} 
\end{equation}
describes the two-valley Coulomb short-range interaction with $V_{0}>0$ and $V_{1}>0$ parameterizing the intra- and intervalley effectively repulsive interactions~\cite{PhysRevLett.132.186401,PhysRevB.110.045427}, $n_{s \xi}$ is the number of particles with spin $s=\uparrow,\downarrow$ in $K$($\xi=+$) or $K^{\prime}$($\xi=-$) valley, $i$ is the sublattice index.
Here, $\psi_{i,s \xi} (\tau,\mathbf{x} )$ is a fermionic field at a state $(s,\xi)$ on a sublattice site $i$, $\tau\in\left[ 0,\beta\right]$ is the imaginary time, $\beta=1/T$ is the inverse temperature. We define $n_{s \xi}(\tau,\mathbf{x} )$ as follows:
\begin{equation}
n_{s \xi} = \sum_{i} \psi^{\dagger}_{i,s \xi} (\tau,\mathbf{x} ) \psi_{i,s \xi} (\tau,\mathbf{x} ) .
\end{equation}
In general, we may use any $H_0$ as it is used to define the bare Green's function.

The interaction part of the Hamiltonian density (\ref{eq:int-hamiltonian}) may be rewritten in terms of number density $n(\tau,\mathbf{x})$, intravalley spin magnetization densities in each valley $m_{\xi}^{\text{S}}(\tau , \mathbf{x})$, orbital magnetization density $m^{\text{O}}(\tau,\mathbf{x})$, which are defined as follows:
\begin{equation}
\begin{aligned}
n(\tau,\mathbf{x})&=\frac{1}{2}\sum_{s , \xi} \sum_{i} \psi^{\dagger}_{i,s \xi} (\tau,\mathbf{x} ) \psi_{i,s \xi} (\tau,\mathbf{x} ) ,\\
m^{\text{S}}_{\xi}(\tau,\mathbf{x})&=\frac{1}{2}\left[\psi^{\dagger}_{i,\uparrow \xi} (\tau,\mathbf{x} ) \psi_{i,\uparrow \xi} (\tau,\mathbf{x} )  -  \psi^{\dagger}_{i,\downarrow \xi} (\tau,\mathbf{x} ) \psi_{i,\downarrow \xi} (\tau,\mathbf{x} )  \right],\\
m^{\text{O}}(\tau,\mathbf{x})&=\frac{1}{2}\sum_{s}\left[\psi^{\dagger}_{i,s +} (\tau,\mathbf{x} ) \psi_{i,s +} (\tau,\mathbf{x} )  -  \psi^{\dagger}_{i,s -} (\tau,\mathbf{x} ) \psi_{i,s -} (\tau,\mathbf{x} )  \right].
\end{aligned}
\end{equation}
Hence, the interaction Hamiltonian reads as follows:
\begin{equation}
\label{eq:int-hamiltonian-rewritten}
\begin{aligned}
H_{\text{int}}(\tau,\mathbf{x})&= \left(\frac{V_0}{2}+V_1 \right) \left[ n(\tau,\mathbf{x}) \right]^2  - \frac{1}{2}\left( 2V_1 -V_0\right) \left[ m^{\text{O}}(\tau,\mathbf{x})\right]^2-V_0 \left[m^{\text{S}}_{+}(\tau,\mathbf{x}) \right]^2-V_0 \left[m^{\text{S}}_{-}(\tau,\mathbf{x}) \right]^2.
\end{aligned} 
\end{equation}

The partition function is defined as
\begin{equation}
\label{eq:part-func-def}
\begin{aligned}
Z=\int D \psi_{i,s \xi} D \psi_{i,s \xi}^{\dagger} \, e^{-S\left[ \psi_{i,s \xi} , \psi^{\dagger}_{i,s \xi}  \right]}
\end{aligned}
\end{equation}
where the action $S\left[ \psi_{i,s \xi}, \psi^{\dagger}_{i,s \xi}  \right]$ is given by
\begin{equation}
S\left[ \psi_{i,s \xi}, \psi^{\dagger}_{i,s \xi}  \right] = \int_{0}^{\beta} d\tau \int d \mathbf{x} \left[ \sum_{s,\xi} \sum_{i}\psi^{\dagger}_{i,s \xi} (\tau ,\mathbf{x}) \partial_{\tau} \psi_{i,s \xi} (\tau ,\mathbf{x})  + H(\tau ,\mathbf{x})\right] .
\end{equation}

We expect that the charge density fluctuations can be neglected, therefore, we focus only on $m^{\text{S}}_{\xi}(\tau,\mathbf{x})$ and $m^{\text{O}}(\tau,\mathbf{x})$. By performing the Hubbard-Stratonovich transformation~\cite{PhysRevB.14.1165,PhysRevB.48.7183,e2002-00356-9}, we rewrite the partition function (\ref{eq:part-func-def}) in the following form:
\begin{equation}
\label{eq:part-func-full}
Z=\int D\Phi_{\text{O}} D\Phi_{\text{S},\,\xi} \, e^{-\int_{0}^{\beta} d\tau \int d\mathbf{x} \left[ \Phi_{\text{O}}^2(\tau,\mathbf{x}) / 2\left(2V_1-V_0\right) +\sum_{\xi} \Phi_{\text{S},\,\xi}^2(\tau,\mathbf{x})/ 2 V_0  \right] } Z_0 \left[ \Phi_{\text{O}} , \Phi_{\text{S},\,\xi}\right] 
\end{equation}
by introducing the Hubbard-Stratonovich fields $\Phi_{\text{O}}$, $\Phi_{\text{S},\,\xi}$ corresponding to $m^{\text{S}}_{\xi}(\tau,\mathbf{x})$ and $m^{\text{O}}(\tau,\mathbf{x})$.
Here, $Z_0 \left[ \Phi_{\text{O}} , \Phi_{\text{S},\,\xi}\right]$ is obtained after integration over the fermionic fields $\psi_{i,s \xi}, \psi_{i,s \xi}^{\dagger}$ and defined as follows:
\begin{equation}
\begin{aligned}
 Z_0 \left[ \Phi_{\text{O}} , \Phi_{\text{S},\,\xi}\right] &=\int D \psi_{i,s \xi} D \psi_{i,s \xi}^{\dagger} \, e^{-\int_{0}^{\beta} d\tau \int d\mathbf{x} \left[\sum_{s,\xi} \sum_{i}\psi^{\dagger}_{i,s \xi} (\tau ,\mathbf{x}) \partial_{\tau} \psi_{i,s \xi} (\tau ,\mathbf{x})  + H_0(\tau ,\mathbf{x}) \right]} \\
&\times e^{-\int_{0}^{\beta} d\tau \int d\mathbf{x} \left[ \Phi_{\text{O}}(\tau,\mathbf{x}) m^{\text{O}}(\tau,\mathbf{x}) +\sum_{\xi}  \sqrt{2} \Phi_{\text{S},\,\xi}(\tau,\mathbf{x}) m^{\text{S}}_{\xi}(\tau,\mathbf{x})  \right]  } .
\end{aligned} 
\end{equation}
Explicitly, $Z_0 \left[ \Phi_{\text{O}} , \Phi_{\text{S},\,\xi}\right]$ is given by
\begin{equation}
\begin{aligned}
 Z_0 \left[ \Phi_{\text{O}} , \Phi_{\text{S},\,\xi}\right] =\text{exp} \left\{ \sum_{s,\xi} \text{Tr} \, \log \left[ I - \left( \frac{\xi}{2}  \hat{\Phi}_{\text{O}}+\frac{s}{2} \sqrt{2}\hat{\Phi}_{\text{S},\,\xi} \right) \hat{G}^{(0)}\right]\right\} ,
\end{aligned} 
\end{equation}
where $\hat{G}^{(0)}$ is a bare Green's function
\begin{equation}
\left[ \partial_{\tau} + H_0(\tau ,\mathbf{x})\right] \hat{G}^{(0)} =-\delta(\tau-\tau^{\prime}) \delta(\mathbf{x}-\mathbf{x}^{\prime})
\end{equation}
and the matrix $\hat{\Phi}$ has the structure $\hat{\Phi}=\Phi(\tau,\mathbf{x}) \delta_{\mathbf{x} \mathbf{x}^{\prime}} \delta(\tau-\tau^{\prime})$. 

In the mean-field approximation, we consider the Hubbard-Stratonovich fields as constants, therefore, using a Fourier-transformed Green's function, we obtain
\begin{equation}
\log Z_0 \left[ \Phi_{\text{O}} , \Phi_{\text{S},\,\xi}\right]  =  \sum_{s,\xi}  \log  \text{det} \left[ I - \left( \frac{\xi}{2}  \hat{\Phi}_{\text{O}}+\frac{s}{2} \sqrt{2}\hat{\Phi}_{\text{S},\,\xi} \right) \hat{G}^{(0)}\right]  .
\end{equation}
From Eq.~(\ref{eq:part-func-full}), we obtain the Landau free energy $\mathcal{F}(\Phi)$
\begin{equation}
\mathcal{F}(\Phi) =  \mathcal{N} \left[ \frac{ \Phi_{\text{O}}^2}{ 2\left(2V_1-V_0\right)} +\sum_{\xi} \frac{\Phi_{\text{S},\,\xi}^2}{ 2 V_0 }  \right] -\frac{1}{\beta}\log Z_0 \left[ \Phi_{\text{O}} , \Phi_{\text{S},\,\xi}\right] . 
\end{equation}
By expanding $\log Z_0$ up to $\hat{\Phi}^{4}$, we obtain the following approximate expressions for the Landau free energy:
\begin{equation}
\log Z_0\left[ \Phi_{\text{O}} , \Phi_{\text{S},\,\xi}\right] \simeq-\beta\mathcal{F}^{(2)}-\beta\mathcal{F}^{(3)}-\beta\mathcal{F}^{(4)} ,\\
\end{equation}
where
\begin{equation}
\begin{aligned}
\mathcal{F}^{(2)} &=- \frac{ \mathcal{N} }{2}\left( \Phi_{\text{O}}^2 + \sum_{\xi} \Phi_{\text{S},\,\xi}^2 \right) \left\{ - \frac{1}{\beta \mathcal{N}}  \sum_{\omega_n}\sum_{\mathbf{k}} \text{tr} \left[ G_{\xi}^2(i \omega_n;\mathbf{k}) \right] \right\} ,\\
\end{aligned} 
\end{equation}
\begin{equation}
\begin{aligned}
\mathcal{F}^{(3)}&=-\frac{ \mathcal{N} }{2} \sum_{\xi} \xi \Phi_{\text{O}}\Phi_{\text{S},\,\xi}^2 \Bigg\{ -\frac{1}{\beta \mathcal{N}} \sum_{\omega_n}\sum_{\mathbf{k}} \text{tr} \left[ G_{\xi}^3(i \omega_n;\mathbf{k} )\right] \Bigg\} ,\\
\end{aligned} 
\end{equation}
\begin{equation}
\begin{aligned}
\mathcal{F}^{(4)}&=- \frac{ \mathcal{N}}{16 } \left(   \Phi_{\text{O}}^4+ \sum_{\xi=\pm} 2   \Phi_{\text{S},\,\xi}^4+ \sum_{\xi=\pm} 6   \Phi_{\text{O}}^2 \Phi_{\text{S},\,\xi}^2 \right) \Bigg\{ -\frac{1}{\beta \mathcal{N}}  \sum_{\omega_n}\sum_{\mathbf{k}} \text{tr} \left[ G_{\xi}^4(i \omega_n; \mathbf{k}) \right] \Bigg\} ,\\
\end{aligned} 
\end{equation}
with $\omega_{n}=\left(2n+1\right)\pi T$ being the fermionic Matsubara frequency and $\mathcal{N}$ being the number of $\mathbf{k}$-points in the summation.
Therefore, we obtain the following Landau free energy density $f(\Phi)=\mathcal{F}(\Phi)  / \mathcal{N}$:
\begin{equation}
\begin{aligned}
f(\Phi)&= a_1  \Phi_{\text{O}}^2 +a_2 \sum_{\xi=\pm}  \Phi_{\text{S},\,\xi}^2+b  \sum_{\xi=\pm} \xi  \Phi_{\text{O}} \Phi_{\text{S},\,\xi}^2 +c \left( \ \Phi_{\text{O}}^4 + \sum_{\xi=\pm} 2    \Phi_{\text{S},\,\xi}^4+ \sum_{\xi=\pm} 6 \Phi_{\text{O}}^2 \Phi_{\text{S},\,\xi}^2 \right) ,
\end{aligned} 
\end{equation}
where $a_{1,2}$, $b$ and $c$ are defined as:
\begin{equation}
\label{eq:landau-free-energy-coefficients-def}
\begin{aligned}
a_1&=\frac{1}{2 \left(2V_1-V_0\right)} \left\{ 1+\left(2V_1-V_0\right)\frac{1}{\beta \mathcal{N}}  \sum_{\omega_n}\sum_{\mathbf{k}} \text{tr} \left[ G_{\xi}^2(i \omega_n;\mathbf{k})  \right] \right\} ,\\
a_2&=\frac{1}{2 V_0} \left\{ 1+V_0 \frac{1}{\beta \mathcal{N}}  \sum_{\omega_n}\sum_{\mathbf{k}} \text{tr} \left[ G_{\xi}^2(i \omega_n;\mathbf{k})  \right] \right\} ,\\
b&=-\frac{1}{ 2 \beta \mathcal{N} }  \sum_{\omega_n}\sum_{\mathbf{k}} \text{tr} \left[ G_{\xi}^3(i \omega_n;\mathbf{k}) \right] ,\\
c&=-\frac{1}{ 16 \beta \mathcal{N} } \sum_{\omega_n}\sum_{\mathbf{k}} \text{tr} \left[ G_{\xi}^4(i \omega_n;\mathbf{k}) \right] .\\
\end{aligned} 
\end{equation}

From the condition for the extrema of $f(\Phi)$, 
\begin{equation}
\frac{\partial f}{\partial  \Phi_{\text{S},\,\xi}}=0,\quad
 \frac{\partial f}{\partial  \Phi_{\text{O}}}=0,\quad
\Phi_{\text{O}},\Phi_{\text{S},\,\xi} \in \mathbb{R},\quad
\end{equation}
we obtain the system of equations for determining the behavior of the order parameters with the chemical potential $\mu$ and temperature $T$:
\begin{equation}
\label{eq:main-order-parameter}
\begin{aligned}
\Phi_{\text{S},\,\xi} \left( a_2 +4 c \Phi_{\text{S},\,\xi}^2+6 c \Phi_{\text{O}}^2+b  \xi \Phi_{\text{O}} \right)&=0 ,\\
\Phi_{\text{O}} \left( a_1 +2 c \Phi_{\text{O}}^2+6 c  \sum_{\xi=\pm}   \Phi_{\text{S},\,\xi}^2 \right)+\frac{b}{2 }   \sum_{\xi=\pm} \xi \Phi_{\text{S},\,\xi}^2 &=0 .\\
\end{aligned} 
\end{equation}
Here, we observe that spin magnetization order parameters are coupled to the orbital magnetization order parameter non-equivalently by the cubic term, which leads to different types of order-mixing. Let us analyze Eq.~(\ref{eq:main-order-parameter}) in the case of $SU(4)$-spin symmetric interaction, i.e. $V_0=V_1=V$.

From Eq.~(\ref{eq:landau-free-energy-coefficients-def}), the coefficients $a_{1,2}$, $b$ and $c$ are expressed in terms of bare Lindhard susceptibility
\begin{equation}
\chi_0(\mu,T)=- \frac{1}{\beta \mathcal{N}}  \sum_{\omega_n}\sum_{\mathbf{k}} \text{tr} \left[ G_{\xi}^2(i \omega_n;\mathbf{k}) \right] 
\end{equation}
as follows
\begin{equation}
\begin{aligned}
a_1&=\frac{1-\left(2V_1-V_0\right)\chi_0}{2\left(2V_1-V_0\right)},\quad a_2=\frac{1-V_0 \chi_0}{2V_0},\quad b=-\frac{1}{4} \frac{\partial \chi_0}{\partial \mu},\quad c=\frac{1}{96 } \frac{\partial^2 \chi_0}{\partial \mu^2} ,\\
\end{aligned} 
\end{equation}
where we have used the equality
\begin{equation}
\sum_{\omega_n}\sum_{\mathbf{k}} \text{tr} \left[ G_{\xi}^{m+2} (i \omega_n;\mathbf{k}) \right] = \frac{(-1)^{m}}{(m+1)!} \frac{\partial^{m}}{\partial \mu^{m}}\sum_{\omega_n}\sum_{\mathbf{k}} \text{tr} \left[ G_{\xi}^{2} (i \omega_n;\mathbf{k}) \right],\,\,m \in \mathbb{N} .
\end{equation}

For $SU(4)$-spin symmetric interaction, i.e. $V_0=V_1=V$, we have $a_1=a_2= a $. Therefore, we solve Eq.~(\ref{eq:main-order-parameter}) for determining the values of the order parameters that minimize $f(\Phi)$.

We model the orbital physics of R6G by a realistic hopping Hamiltonian~\cite{PhysRevB.82.035409,
koshino_trigonal_2009,}, which in momentum space reads

\begin{align}
    \mathcal{H}_0(\mathbf{k};s,\xi)=&
    \begin{bmatrix}
    Dz_l& \gamma_0f(\mathbf{k})\\
    \gamma_0f^*(\mathbf{k})& Dz_l
    \end{bmatrix}\delta_{l,l'}
    +
    \begin{bmatrix}
    \gamma_4f^{*}(\mathbf{k})& \gamma_1\\
    \gamma_3f(\mathbf{k})&  \gamma_4f^{*}(\mathbf{k})
    \end{bmatrix}\delta_{l,l'+1}
    +
    \begin{bmatrix}
    \gamma_4f(\mathbf{k})& \gamma_3f^*(\mathbf{k})\\
    \gamma_1&  \gamma_4f(\mathbf{k})
    \end{bmatrix}\delta_{l+1,l'}
    \nonumber
    \\
    +&\begin{bmatrix}
    0& 0\\
    \gamma_6&0
    \end{bmatrix}\delta_{l,l'+2}
    +
    \begin{bmatrix}
    0& \gamma_6\\
    0&0
    \end{bmatrix}\delta_{l+2,l'}
    \label{eq:tb}
\end{align}
acting on single-particle Bloch states with momenta $\mathbf{k}=(k_x, k_y)$---measured 
from $K$ and $K'$ valleys---that are based on carbon $p_z$ orbitals residing on the R6G 
sublattices ($A_l, B_l$), where $l$ is the layer index.  
Here, $f(\mathbf{k})=-\left(\sqrt{3}a/2\right)\left(\xi k_x-ik_y\right)$ is the linearized nearest-neighbor structure factor, $a$ is graphene's lattice constant, and $\xi=\pm$ is the valley index; ${\gamma_i}$ are orbital hopping parameters taken from ~\cite{PhysRevB.82.035409,
koshino_trigonal_2009}. 
 The electrostatic potentials $D$ on different layers are incorporated into Eq.~(\ref{eq:tb}) through the different on-site energies $Dz_l$, where $z_l$ is $z$ coordinate of $l$-layer.

The Lindhard's susceptibility $\chi_{0}$ can be given in terms of eigenenergies $\varepsilon_{n\textbf{k}}$ and eigenvectors $u_{n\textbf{k}}^{ai}$ of the Hamiltonian $H_0$ as follows:
\begin{align}
    \chi_{0}=\frac{A_{uc}}{2}\sum_{ab}\int_{|\textbf{k}|<\Lambda}\frac{\mathrm{d}^2\textbf{k}}{(2\pi)^2}\,\sum_{nm}\sum_{ij}
    \frac{f_{n\textbf{k}}-f_{m\textbf{k}}}{\varepsilon_{m\textbf{k}}-\varepsilon_{n\textbf{k}}}
    (u_{n\textbf{k}}^{ai})^{*}
    u_{m\textbf{k}}^{bi}
    (u_{m\textbf{k}}^{bj})^{*}
    u_{n\textbf{k}}^{aj}\,,
\end{align}
where $A_{uc}$ stand s for the unit cell area and the corresponding Fermi-Dirac weights
$f_{n\textbf{k}}$ (we suppressed the valley index) read:
\begin{equation}
f_{n\textbf{k}}=\frac{1}{1+\exp\left[\left( \varepsilon_{n\textbf{k}} - \mu \right)/T \right]}\,.
\end{equation}

The static susceptibility $\chi_{0}$ is computed at $T = 1$~K and the momentum cutoff $\Lambda= 0.06~\AA^{-1}$. 
The phase diagram is determined by solutions for which the free energy $f(\Phi)$ from \eqref{eq:landau} has the global minimum. 

\begin{figure*}[t]
     \centering
     \includegraphics[width=1\linewidth]{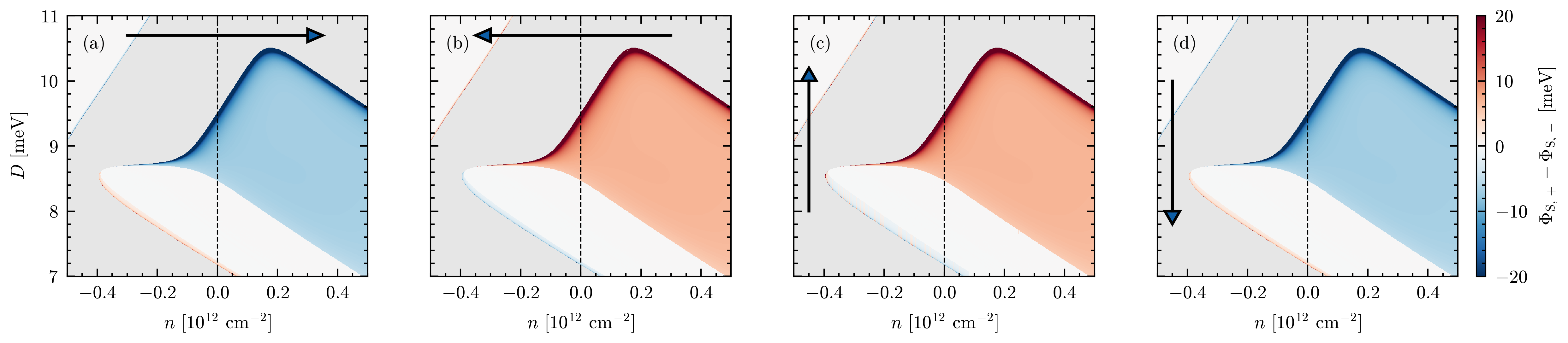} 
     \caption{\textbf{a--d} Difference of the spin magnetization order parameters $\Phi_{\text{S},\,+}-\Phi_{\text{S},\,-}$ as a function of doping $n$ and displacement field $D$. The black arrow on the panels indicates the direction of the parameter sweep.}
     \label{fig: supp8}
\end{figure*}

\begin{figure*}[t]
    \centering
     \includegraphics[width=1\linewidth]{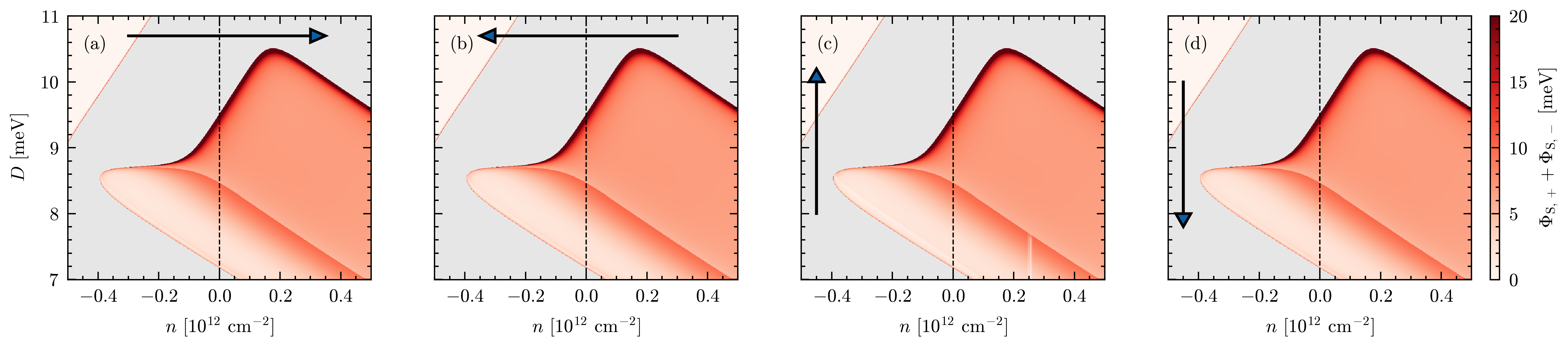} 
     \caption{\textbf{a--d} Sum of the spin magnetization order parameters $\Phi_{\text{S},\,+}+\Phi_{\text{S},\,-}$ as a function of doping $n$ and displacement field $D$. In the region, where $\Phi_{\text{S},\,+}=\Phi_{\text{S},\,-}$ (compare with Extended Data Figure~\ref{fig: supp8}), the system has only spin magnetization order without any orbital magnetization. The black arrow on the panels indicates the direction of the parameter sweep.}
     \label{fig: supp9}
\end{figure*}

In addition to phase diagram of the orbital magnetization presented in Figure~\ref{fig: figure2}\textbf{e}-\textbf{h}, we calculate the phase diagrams for the spin magnetization order parameter. Since Landau free energy (\ref{eq:landau}) depends on $\Phi_{\text{S},\,\xi}^2$ and $\Phi_{\text{S},\,\xi}^4$, we set the order parameter $\Phi_{\text{S},\,\xi}$ as nonnegative. Therefore, we present the difference $\Phi_{\text{S},\,+}-\Phi_{\text{S},\,-}$ and the sum $\Phi_{\text{S},\,+}+\Phi_{\text{S},\,-}$ as a function of $(n,D)$ in Extended Data Figures~\ref{fig: supp8} and \ref{fig: supp9} respectively. We obtain that the spin magnetization dominates in a particular valley depending on the sign of the orbital magnetization order parameter $\Phi_{\text{O}}$; compare Figure~\ref{fig: figure2}\textbf{e}-\textbf{h} with Extended Data Figure~\ref{fig: supp8}\textbf{a}-\textbf{d}. Therefore, we obtain different types of correlated order mixing depending on the direction of the parameter sweeping. In the region with vanishing orbital magnetization, the spin magnetization order parameters are equal to each other; see Extended Data Figure~\ref{fig: supp9}.

To stabilize the numerical optimization, we also introduced a small term in the Landau free energy~(\ref{eq:landau}):
\begin{equation}
f^{\text{reg}}(\Phi)=\eta\left(\Phi_{\mathrm{S},+}-\Phi_{\mathrm{S},-}\right), \qquad \eta \to 0.
\end{equation}

\end{document}